\documentclass[12pt,preprint]{aastex}

\input{psfig.sty}
\def\etal   {{et~al.\/}}

\def\mo     {{$M_{\odot}$}}

\begin{document}

\shorttitle{$\beta$ Pic Analogs in GLIMPSE}
\shortauthors{Uzpen, B. \etal}

\title{Identification of Main Sequence Stars with Mid-Infrared Excesses
Using
GLIMPSE: $\beta$-Pictoris Analogs? }

\author{B. Uzpen, \altaffilmark{1} H.A. Kobulnicky, \altaffilmark{1}
K. A. G. Olsen, \altaffilmark{2} D. P. Clemens,\altaffilmark{3} T.L. Laurance,
\altaffilmark{1} M. R. Meade,\altaffilmark{4}
B. L. Babler,\altaffilmark{4} R. Indebetouw,\altaffilmark{4}
B. A. Whitney,\altaffilmark{5} C. Watson,\altaffilmark{4}
M. G. Wolfire,\altaffilmark{6} M. J. Wolff,\altaffilmark{5}
R. A. Benjamin,\altaffilmark{7} T. M. Bania,\altaffilmark{3}
M. Cohen,\altaffilmark{8} K. E. Devine,\altaffilmark{4}
J. M. Dickey,\altaffilmark{9} F. Heitsch,\altaffilmark{10}
J. M. Jackson,\altaffilmark{3} A. P. Marston,\altaffilmark{11}
J. S. Mathis,\altaffilmark{4} E. P. Mercer,\altaffilmark{3}
J. R. Stauffer,\altaffilmark{12} S. R. Stolovy,\altaffilmark{12}
D.E. Backman, \altaffilmark{13} \and E. Churchwell \altaffilmark{4} }


\altaffiltext{1}{University of Wyoming, Dept. of Physics \&
Astronomy, Dept. 3905, Laramie, WY 82071}

\altaffiltext{2}{Cerro Telolo Inter-American
Observatory\footnote{Visiting astronomer, Cerro Tololo Inter-American
Observatory, National Optical Astronomy Observatory, which is operated
by the Association of Universities for Research in Astronomy, under
contract with the National Science Foundation.}, National Optical
Astronomy Observatory, Casilla 603, La Serena Chile}

\altaffiltext{3}{Boston University, Institute for Astrophysical
Research, 725 Commonwealth Ave., Boston, MA 02215}

\altaffiltext{4}{University of Wisconsin-Madison, Dept. of Astronomy,
475 N. Charter St., Madison, WI 53706}

\altaffiltext{5}{Space Science Institute, 4750 Walnut St. Suite 205,
Boulder, CO 80301}

\altaffiltext{6}{University of Maryland, Dept. of Astronomy,
College Park, MD 20742-2421}

\altaffiltext{7}{University of Wisconsin-Whitewater, Physics Dept.,
800 W. Main St., Whitewater, WI 53190}

\altaffiltext{8}{University of California-Berkeley, Radio Astronomy
Lab, 601 Campbell Hall, Berkeley, CA 94720}

\altaffiltext{9}{University of Minnesota, Dept. of Astronomy, 116
Church St., SE, Minneapolis, MN 55455}

\altaffiltext{10}{Institute for Astronomy \& Astrophysics, University of Munich,
Scheinerstrasse 1, 81679 Munich
}

\altaffiltext{11}{ESTEC/SCI-SA,Postbus 299,2200 AG Noordwijk,The
Netherlands}

\altaffiltext{12}{Caltech, Spitzer Science Center, MS 314-6, Pasadena,
CA 91125}

\altaffiltext{13}{SOFIA, MS 211-3, NASA-Ames Research Center, Moffett Field, CA
94035-1000}





\vskip 1.cm

\begin{abstract}

\textit{Spitzer} IRAC 3.6-8 $\mu$m photometry obtained as part of the
GLIMPSE survey has revealed mid-infrared excesses for $33$ field stars
with known spectral types in a $1.2$ sq. degree field centered on the
southern Galactic \ion{H}{2} region RCW$49$. These stars comprise a
subset of 184 stars with known spectral classification, most of which
were pre-selected to have unusually red IR colors. We propose that the
mid-IR excesses are caused by circumstellar dust disks that are either
very late remnants of stellar formation or debris disks generated by
planet formation. Of these 33 stars, 29 appear to be main-sequence
stars based on optical spectral classifications.  Five of the 29
main-sequence stars are O or B stars with excesses that can be
plausibly explained by thermal bremsstrahlung emission, and four are
post main-sequence stars. The lone O star is an O4V((f)) at a
spectrophotometric distance of 3233$^{+ 540}_{- 535}$ pc and may be
the earliest member of the Westerlund 2 cluster. Of the remaining 24
main-sequence stars, 18 have SEDs that are consistent with hot dusty
debris disks, a possible signature of planet formation. Modeling the
excesses as blackbodies demonstrates that the blackbody components
have fractional bolometric disk-to-star luminosity ratios,
$\frac{L_{IR}}{L_{*}}$, ranging from 10$^{-3}$ to 10$^{-2}$ with
temperatures ranging from 220 to 820 K. The inferred temperatures are
more consistent with asteroid belts rather than the cooler
temperatures expected for Kuiper belts.  Mid-IR excesses are found in
all spectral types from late B to early K.

\end{abstract}


\section{Introduction}

IRAS observations of Vega detected an infrared excess above the
photospheric level, suggesting the presence of circumstellar material
(Aumann \etal \ 1984). Follow-up observations found infrared excesses
around other main-sequence stars, including $\beta$
Pictoris. Coronographic imaging identified a resolved dusty disk
around $\beta$ Pictoris as the cause of the infrared excess (Smith \&
Terrile 1984). $\beta$ Pictoris's disk is a prototype for a young, but
nearly complete planetary system. Identification of this disk was the
first observational evidence of planetary systems outside our
own. Many nearby stars have been investigated using IRAS and ISO to
search for far-IR photospheric excesses, and a large number have been
found (Lagrange \etal \ 2000; Backman \etal \ 1993; Chen \etal \ 2005
and references therein). Due to the large beam size of IRAS, far-IR
observations of nearby stars such as $\beta$ Pictoris, $\epsilon$
Eradani, Fomalhaut, and others have resulted in follow-up higher
resolution ground based mid-IR observations that resolved their disks
and probed their compositions (Lagrange \etal \ 2000). Stars with
far-IR excesses such as $\epsilon$ Eradani may even harbor planets
(Hatzes \etal \ 2000).

Only a few main-sequence stars with mid-IR but no near-IR excesses
have been detected. Aumann \& Probst (1991) investigated nearby
main-sequence stars using 12 $\mu$m measurements from IRAS, but out of
the $548$ stars investigated, only $13$ stars had colors unusual
enough to warrant further investigation. Follow up ground-based
observations were conducted on $7$ stars at $10$ $\mu$m to determine
if the excess originated from the star. Only $\beta$ Pictoris and
$\zeta$ Leporis exhibited mid-IR excesses (Aumann \& Probst
1991). Chen \& Jura (2001) measured a temperature of 370 K for the
mid-IR excess in $\zeta$ Leporis and suggest that it is due to a
massive asteroid belt.  The excess in $\zeta$ Leporis originates at
small radii, less than 6 AU, and not at large radii, $\geq$40 AU,
unlike other stars such as $\beta$ Pictoris which have large disks
extending hundreds of AU (Chen \& Jura 2001). Stars with higher disk
temperatures have dust which is closer to the star, and the excess may
be more analogous to the warm temperatures of asteroid belts than to
the $\sim100$K dust of Kuiper belts.

Evidence for planets around nearby stars is now overwhelming with over
$120$ planets in $100+$ planetary systems\footnote{Extrasolar Planet
Encyclopedia http://cfa-www.harvard.edu/planets}, most found using
high resolution stellar spectroscopy capable of detecting Doppler
shifts in the parent star. However, the process or processes by which
these planets form is still unknown. Planetesimals may originate in
circumstellar disks, formed as a by-product of star formation. The
evolutionary sequence of both inner and outer circumstellar disks from
pre-main sequence to main-sequence is not well understood (Meyer \&
Beckwith 2000). Lack of observational evidence of an evolutionary disk
sequence is a major hindrance in the understanding of planetary
formation. The reason for this lack of evidence is that only the
nearest stars have been probed for circumstellar disks, and a
relatively small number of stars with disks have been identified. It
is estimated that at least 15\% of nearby A-K main-sequence stars
have dusty debris disks that were detectable to IRAS and ISO
sensitivities in the far-infrared (Lagrange \etal \ 2000; Backman \&
Paresce 1993).

The \textit{Spitzer} Legacy program includes two surveys
investigating the formation of stars and the processes leading to
the formation of solar systems: the ``Cores to Disks'' (C2D; Evans
\etal \ 2003) and ``Formation and Evolution of Planetary Systems''
(FEPS; Meyers \etal \ 2001) projects. Data from these programs
will cover a large portion of the spectral energy distribution
(SED) of stars as well as provide insight into the structure and
evolution of circumstellar disks.  The focus of the C2D project is
on stellar cores and very young stars with ages up to $\sim$$10$
Myr. The FEPS project focuses on older solar analogues, 3 Myr to 3
Gyr, to determine disk composition and characterize their
evolutionary stages. The FEPS and C2D projects are, however,
narrowly focused in spatial coverage, with C2D covering $\sim$$20$
sq. degree and FEPS focusing on $\sim$$350$ individual stars out to
$\sim$$200$ pc.

In this present study, we have utilized a subset of the Galactic
Legacy Infrared MidPlane Survey Extraordinaire (GLIMPSE; Benjamin et
al. 2003) database to try to identify main-sequence stars with
mid-infrared excesses. Given the sensitivity of the \textit{Spitzer
Space Telescope} Spitzer, we expect to be able to detect unreddened A
stars to $\sim$$4$ kpc, F stars to $\sim$$2$ kpc, G stars to
$\sim$$1.3$ kpc, and K stars to $\sim$$1$ kpc. Given these distances
we will survey a much larger volume of space for stars exhibiting
mid-IR excesses than any previous study.

Optical spectra were obtained to determine if near and mid-infrared
colors could indicate a spectral sequence.  In this paper, we present
results of a survey of the GLIMPSE Observation Strategy Validation
(OSV) region in the vicinity of the southern Galactic star-forming
region RCW $49$, where the Westerlund 2 open cluster is centered
(Churchwell \etal \ 2004). We obtained optical spectroscopy for a
sample of stars with red or unusual near and mid-infrared colors. We
combine optical photometry and spectroscopy with near and mid-infrared
photometry to model the SEDs of the stars. Once a star is shown to
exhibit a mid-IR excess, we model the excess as a single blackbody to
gain some insight into its basic properties. In Section 2, we discuss
the photometry obtained with GLIMPSE and the method of classification
of spectral types. In Section 3, we discuss how stars with mid-IR
excesses were identified, and in Section 4 we compare the stars with
mid-IR excesses to the prototype debris disk system, $\beta$
Pictoris. In Section 5 we discuss the lone O4V((f)) star which
exhibits a mid-IR excess and its implications for the age and distance
of Westerlund 2. This is a preliminary investigation to determine if
warm debris disks can detected with the GLIMPSE survey in the mid-IR.

\section{Data}

\subsection{GLIMPSE Photometry}

The GLIMPSE project is one of six Spitzer Legacy Programs. GLIMPSE
mapped the Galactic Plane in four infrared array camera (IRAC; Fazio
\etal \ 2004) band passes, $[3.6]$, $[4.5]$, $[5.8]$, and $[8.0]$
$\mu$m from $|l|=10^\circ-65^\circ$ and $|b|<1^\circ$ degrees
(Benjamin \etal \ 2003). This survey has generated a point source
catalog of 3$\times$$10^{7}$ objects. The GLIMPSE program is the
largest, most sensitive, mid-infrared survey to date. The GLIMPSE
program mapped RCW$49$ as part of the OSV (Churchwell \etal \
2004). This region was mapped $10$ times with $1.2$ second exposures
by IRAC and is therefore the deepest region surveyed by the GLIMPSE
program. A total of 38,734 mid-infrared catalog sources were found in
the OSV.  To be included in the Point Source Catalog, a source must
have a signal-to-noise ratio of greater than 5:1 with at least two
detections in one band and at least one detection in an adjacent band
with fluxes greater than 0.6 mJy, 0.6 mJy, 2 mJy, and 10 mJy (in Bands
1 through 4, respectively). The other two bands need only have a
signal-to-noise ratio of greater than 3:1. See the GLIMPSE data
document\footnote{GLIMPSE Team Webpage
http://ssc.spitzer.caltech.edu/legacy} or Mercer \etal \ (2004),
Churchwell \etal \ (2004), Whitney \etal \ (2004), or Indebetouw \etal
\ (2005) for further descriptions.

\subsection{Spectra}

We obtained optical spectroscopy of $220$ stars in the GLIMPSE OSV
field surrounding RCW$49$ using the Hydra instrument on the CTIO $4$m
on the night of $2004$ March $1$ using three fiber configurations. The
instrument configuration used the large $2$ arcsecond fibers and the
KGPL1 grating, which provides a dispersion of $0.6$ \AA/pixel, a
spectral resolution of $4.2$ \AA,\ and wavelength coverage of
$\sim$$2400$ \AA \ from $\sim$$3600-6000$ \AA. This wavelength regime
covers the majority of the Hydrogen Balmer series, Ca H and K, as well
as many metal lines which are necessary for luminosity and temperature
classification. Total exposure times were $20$ minutes for the first
fiber configuration and $60$ minutes for the second and third fiber
configurations. Sky conditions were photometric.

\begin{figure}
    \psfig{file=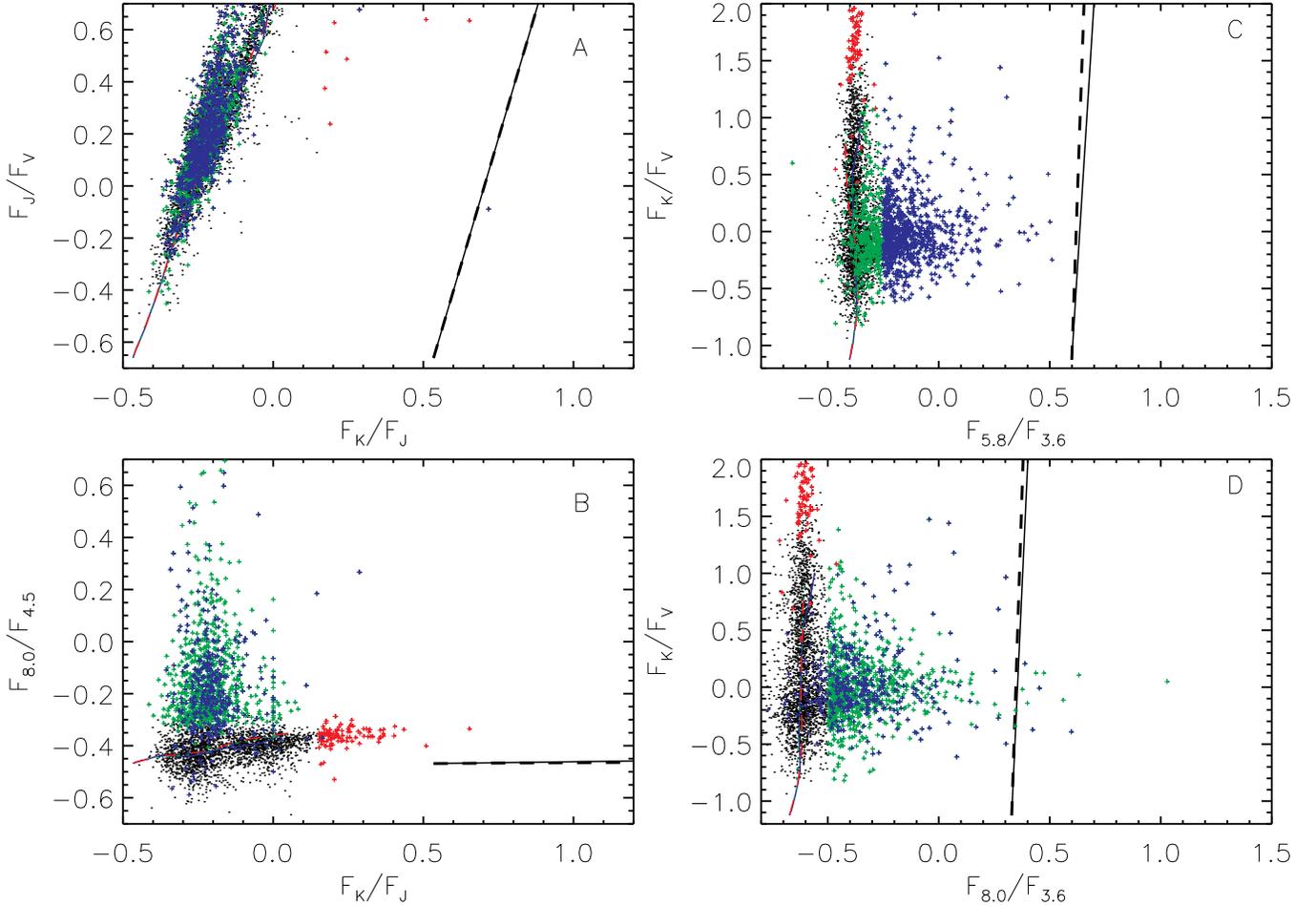}
    \caption{Color-color plots of cataloged point sources in the
    GLIMPSE OSV region (Westerlund 2 region). The black lines are the
    extinction vectors of Li \& Draine (2001). The dashed vectors are
    the extinction laws of Indebetouw (2005) in the mid-IR.  Both sets
    of extinction lines represent $A_V=10$ and are offset from the
    stellar loci for clarity. The green pluses are log
    $\frac{F_{8.0}}{F_{3.6}}$ $\geq$ -0.5 (criterion 1), the red are
    log $\frac{F_{K}}{F_{J}}$ $\geq$ 0.15 (criterion 2), and the blue
    are log $\frac{F_{5.8}}{F_{3.6}}$ $\geq$ -0.25 (criterion 3).}
    \label{color}
\end{figure}

We selected target stars based on preliminary GLIMPSE photometry in
all $4$ IRAC bandpasses along with $2$MASS\footnote{This publication
makes use of data products from the Two Micron All Sky Survey, which
is a joint project of the University of Massachusetts and the Infrared
Processing and Analysis Center/California Institute of Technology,
funded by the National Aeronautics and Space Administration and the
National Science Foundation.}, and Guide Star Catalog $2.2$
(GSC\footnote{The Guide Star Catalogue-II is a joint project of the
Space Telescope Science Institute and the Osservatorio Astronomico di
Torino.  Space Telescope Science Institute is operated by the
Association of Universities for Research in Astronomy, for the
National Aeronautics and Space Administration under contract
NAS5-26555.  The participation of the Osservatorio Astronomico di
Torino is supported by the Italian Council for Research in Astronomy.
Additional support is provided by European Southern Observatory, Space
Telescope European Coordinating Facility, the International GEMINI
project and the European Space Agency Astrophysics Division. This
research has made use of the VizieR catalogue access tool, CDS,
Strasbourg, France.}) photometry. The stars were selected using
optical brightness and infrared colors. The stars in the first fiber
configuration consisted of High-Precision Parallax-Collecting
Satellite (Hipparcos) and Tycho Catalog stars and bright GSC stars
with V $\leq$ 13. Bright stars with $2$MASS/IRAC colors indicative of
possible IR excesses were also placed in the first two fiber
fields. We selected stars with colors indicating an IR-excess or
extreme reddening as those which satisfied one or more of the
following criteria: flux ratios log $\frac{F_{8.0}}{F_{3.6}}$ $\geq$
-0.5 (criterion 1), log $\frac{F_{K}}{F_{J}}$ $\geq$ 0.15 (criterion
2), or log $\frac{F_{5.8}}{F_{3.6}}$ $\geq$ -0.25 (criterion 3). We
chose these criteria to sample a representative distribution of
objects with unusual infrared colors. Criterion 1 sources comprise the
reddest $\sim$$15$ $\%$ of sources with visual counterparts, criterion
2 sources comprise $\sim$$2$ $\%$, and criterion 3 comprise $\sim$$13$
$\%$. Color-color plots of the data set used in selecting stars are
shown in Figure~\ref{color}. Stars that met criterion $1$, $2$, and
$3$ are plotted as green, red, and blue pluses respectively.
Figure~\ref{color} gives a representative view of the color-color
space occupied by the stars meeting one or more of the three criteria.

The second fiber field consisted of stars with GSC photometry and
IR colors satisfying one or more of the three color criteria. The
third fiber field consisted of stars meeting one or more of the
color criteria, but some stars did not have GSC V-band
counterparts, allowing for the inclusion of extremely red stars.

A total of $164$ new stellar classifications resulted from these
observations. The stars were classified by eye using comparison
spectra (Yamashita \etal \ 1976) to within two comparison temperature
classes of the atlas. There were $56$ stars too faint or featureless
for classification. Protostars may comprise a significant fraction of
these unclassified objects because they have high extinctions and are
optically faint. There are $\sim$$50$ stars in the OSV that had
previous spectral classifications in the literature, but most of them
were saturated in GLIMPSE so their mid-IR colors are, therefore,
unreliable and are not included in this investigation. We were able to
use 20 stars that had suitable GLIMPSE colors and were previously
classified in the literature.

\section{Analysis}

The positions of 38,734 GLIMPSE Point Source Catalog objects in the
OSV were matched against the positions of $2$MASS Point Source Catalog
objects using the GLIMPSE data reduction pipeline. We then matched the
Guide Stars with the GLIMPSE objects to produce V, J, H, K, [3.6],
[4.5], [5.8], [8.0] photometry for the 184 spectrally classified
stars, although some objects lack photometry at one or more of the
eight bandpasses. The GSC V-band photometry was extracted from
photometric plates, and most stars have V-band uncertainties of
$\sim$$0.4$ magnitudes, (see the GSC documentation for more
information).

The eight photometric data points were fit with Kurucz ATLAS9 models
using temperatures and effective gravities corresponding to the
nearest spectral type (Kurucz 1993).  Since all of the stars lie
within the solar circle, we adopt the solar metallicity models.  The
Kurucz model surface fluxes were scaled by a factor, $X$, so that the
model K-band fluxes matched the observed K-band 2MASS photometry.
This scale factor represents $X=R^2/D^2$, the ratio of the stellar
radius squared over the distance to the star squared. Adopting stellar
radii appropriate for the observed spectral types (De Jager, C. \&
Nieuwenhuijzen, H. (1987), Schmidt-Kaler, Th. (1982), Johnson,
H.L. (1966)), we calculated a spectrophotometric distance to each
star.  These distances and their uncertainties estimated by taking
into account the K-band photometric errors added in quadrature with an
additional 15\% uncertainty in the stellar radii are listed in the
tables.  The K-band measurements were used as the basis for the
distance computation because the effects of extinction are nearly
negligible at 2.2 $\mu$m for most of our stars. Our derived $A_V$ are
all $<1.5$ mag implying $A_K<0.15 mag$. The model atmospheres were
reddened with variable amounts of extinction using the Li \& Draine
(2001) extinction curve, except in the mid-infrared ($\lambda>3$
$\mu$m) where the GLIMPSE extinction results were used (Indebetouw
2005).  Distance and extinction were fit iteratively as free
parameters covering a grid of all reasonable values.  The best fit
parameters for each star are determined by the minimum of the $\chi^2$
statistic.  Modeling the star with the next earlier or later spectral
type model available in the Kurucz library (i.e., an A3 with an A0 or
an A5) produced very similar reduced $\chi^2$ values. This allows
small errors in classification to yield similar results. This also
provides a check on classification. If a star is grossly misclassified
the fit is poor and the classification is double checked.

We performed a trial fitting procedure in which the effective
temperatures and gravities of the stellar models were allowed to
vary as free parameters (i.e., the program searched through the
entire library of Kurucz models to find a best fit).  In nearly
all cases, this method produced fits with lower $\chi^2$, but this
approach frequently converged upon temperature and gravity
combinations which were grossly inconsistent with the observed
spectral types.  We found that there was often a degeneracy
between the amount of extinction and the effective temperature,
such that models with high temperatures and low extinctions
yielded similar fits as models with low temperatures and high
extinctions.  We, therefore, chose to fix the effective temperatures
and gravities of the models at value appropriate to the classified
spectral type.

Our data set consisted of $184$ main-sequence, giant, and supergiant
stars with spectral types in the OSV region. This sample included 20
stars with known spectral types from the literature which had reliable
GLIMPSE photometry as well as the 164 sources for which we obtained
new classifications. We calculated ($K-8$) color excesses, E($K-8$),
by taking the differences between the Kurucz photospheric model and
the photometric ($K-8$). We chose the ($K-8$) color because
main-sequence stars should exhibit minimal color differences for these
wavelengths. This method is similar to the method used by Aumann \&
Probst (1991), in which $K-12$ from IRAS was used to identify warm
disk candidates.

\begin{figure}
    \psfig{file=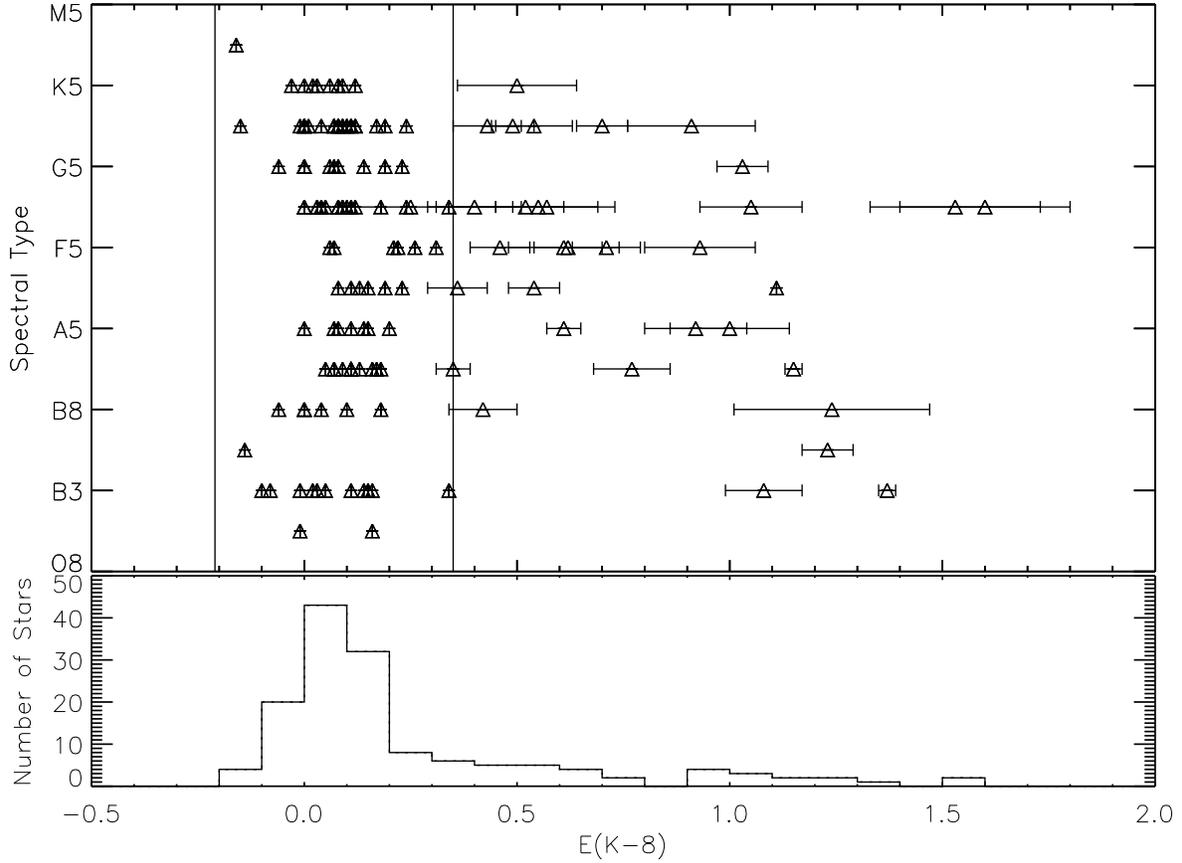,height=4.5in}
\vspace{0.5in}
    \caption{The Kurucz model spectral class versus the ($K-8$) color
excess. The majority of stars have no significant ($K-8$) excess and
cluster between -0.21 and 0.35 (denoted by the vertical bars), while
33 stars have an excess greater than 0.35 mag. The vertical lines
denote the 3 $\sigma$ dispersion from the mean of the 110 stars
without excesses. The distribution shows that there is a long tail
with excesses exhibited by stars of all spectral classes.}
    \label{ek8}
\end{figure}

Figure~\ref{ek8} shows the E($K-8$) distribution of stars with ($K-8$)
colors. The distribution shows a peak with a mean near E($K-8$)=0.08
and a long tail toward larger excesses. The majority of stars, 110,
have E($K-8$) between -0.2 and 0.35 with a mean of 0.08 and a
dispersion of 0.09. There were 33 stars with excesses larger than
three times the dispersion of the main group, i.e., with an E($K-8$)
$\geq$ 0.35. Of these 33 stars with possible mid-IR excesses, three
did not have measurements at [5.8], and three of the stars had
photometric uncertainties placing them within one standard deviation
from an E($K-8$) of 0.35. In order to place a higher reliability on
our detections we required that the stars have all four IRAC
measurements, and an E($K-8$) $\geq$ 0.35 + 1 $\sigma$. Table $1$
gives the colors and uncertainties for the 18 main sequence stars
satisfying the above criteria, along with their spectral types and
color uncertainties. Hereafter, we refer to this group of stars as
the ''G18'' sample. Table 2 contains the stars that had an excess
greater than 0.35 but did not meet all of our criteria for inclusion
in Table 1. Table 2 include stars which did not have all 4 IRAC
measurements, or stars of early spectral type where the excess could
be explained by thermal bremsstrahlung emission (See 3.1).

\begin{figure}
    \psfig{file=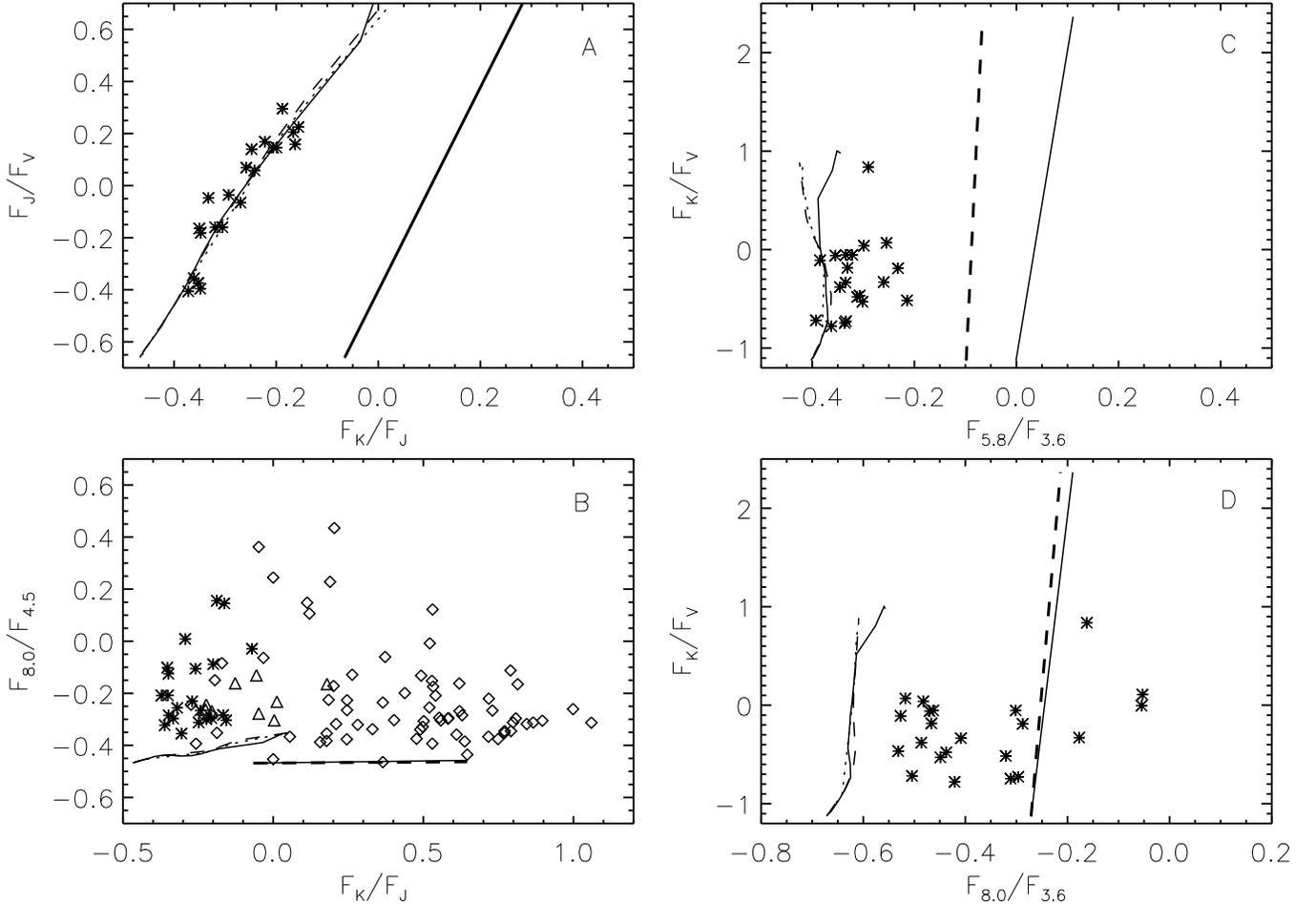}
    \caption{Color-color plots of stars in the GLIMPSE OSV region. The
large black asterisks mark classified stars with a ($K-8$) excess. The
triangles mark classified stars with a ($K-8$) excess which did not
have a V-band GSC measurement. The diamonds denote spectroscopic
target stars which were too faint or featureless for spectral
classification. Most of the diamonds are consistent with being heavily
reddened stars. The black vectors are the extinction vector of Li \&
Draine (2001). The dashed vectors are the extinction law of Indebetouw
(2005) in the mid-IR. Both sets of extinction vectors represent
$A_V=10$ and are offset from the stellar sequence for clarity. The
solid thin line is the theoretical main-sequence, the dash-dot
is the giant, and the dashed is the supergiant sequence.
(De Jager, C. \& Nieuwenhuijzen, H. (1987), Schmidt-Kaler, Th. (1982),
Johnson, H.L. (1966))}
    \label{colore}
\end{figure}

Figure~\ref{colore} shows color-color plots of the stars with mid-IR
excesses. Panel A shows that the stars with a ($K-8$) excess exhibit
typical colors for the J,H,K near-infrared, but in Panel D they show a
significant deviation from the main-sequence which can not be
explained by reddening. Stars which could not be classified because of
insufficient signal to noise in the optical spectra are shown only in
panel B because almost all of these stars were in Field 3 and did not
have a V band measurement. Triangles denote stars with a mid-IR excess
that did not have V-band measurements. Nearly all of the unidentified
stars have $\frac{F_{K}}{F_{J}}$ much redder than the stars exhibiting
mid-infrared excesses. These stars are likely to be highly
extinguished and could be protostars.

\subsection{The Mid-IR Excesses}

We analyzed the SED's of the 23 stars meeting the above ''excess''
criteria to determine whether phenomena other than disks could
explain their mid-IR excesses. An unresolved close stellar
companion cannot produce the observed IR-excess. For most of the
stars in Table $1$, their mid-IR excesses are only significant at
[8.0]. A companion dwarf star of G, K, or M spectral class would
raise the photometric flux at wavelengths $\leq$ $1$ $\mu$m to a
much greater degree than in the mid-IR, and such additional
stellar components are inconsistent with the observed SEDs. Even
with stellar additional companions, the [8.0] measurements are
still above expected photospheric levels. If any of our targets
are multiple star systems, the primary star dominates the SED and
a diskless companion would not contribute significantly to the
SED.

We investigated the GLIMPSE residual images after PSF fitting
photometry and subtraction to rule out the possibility of
systematic photometric errors. There were no significant residuals
in the pixels at or surrounding each star's location. If larger
photometric uncertainties in high background or highly structured
background regions produced spurious signatures of ($K-8$)
excesses, we would expect to see as many stars with negative
E($K-8$) as positive E($K-8$). The distribution in
Figure~\ref{ek8}, however, shows only significant positive
E($K-8$) values. If the excess is due to only PAH features, we
would expect to see an enhancement at [3.6], [5.8], and [8.0] with
respect to [4.5] since [4.5] is devoid of PAH features. None of
our excess sources exhibit this enhancement.

Thermal bremsstrahlung emission could possibly explain the E($K-8$)
values seen in the three early B stars and one O star listed in Table
2. These stars generally have all $4$ IRAC measurements in excess of
the photosphere model values. An optically thin, thermal
bremsstrahlung component was included in the modeling of their
SEDs. The luminosity of the bremsstrahlung component was allowed to
vary as a free parameter along with extinction and distance. These
models with bremsstrahlung components fit the SEDs much better than
the models with a single blackbody component. A blackbody component model
with a single temperature is insufficient to fit all four measurements
in excess of the photosphere. Since these are ionizing stars and they
fit a thermal bremsstrahlung model, we feel that this is the most
likely explanation for the mid-IR excess of these four stars.

In conclusion, we note that the characteristic temperature of the
blackbody component which creates a significant excess at 8
$\mu$m, but not at shorter wavelengths, is $\sim$$400$ K because a
400 K blackbody peaks at 7.4 $\mu$m . Even the coolest known brown
dwarfs (Burgasser et al. 2000; Burrows et al. 2001) have
temperatures of $\sim$$700-750$ K, and Jupiter-like planets would
be insufficiently luminous to create the observed infrared
excesses. We conclude that the source of the mid-IR excess for the
stars in Table 1 is probably warm $\sim$$400$ K dust.

\section{Discussion}

\begin{figure}
    \psfig{file=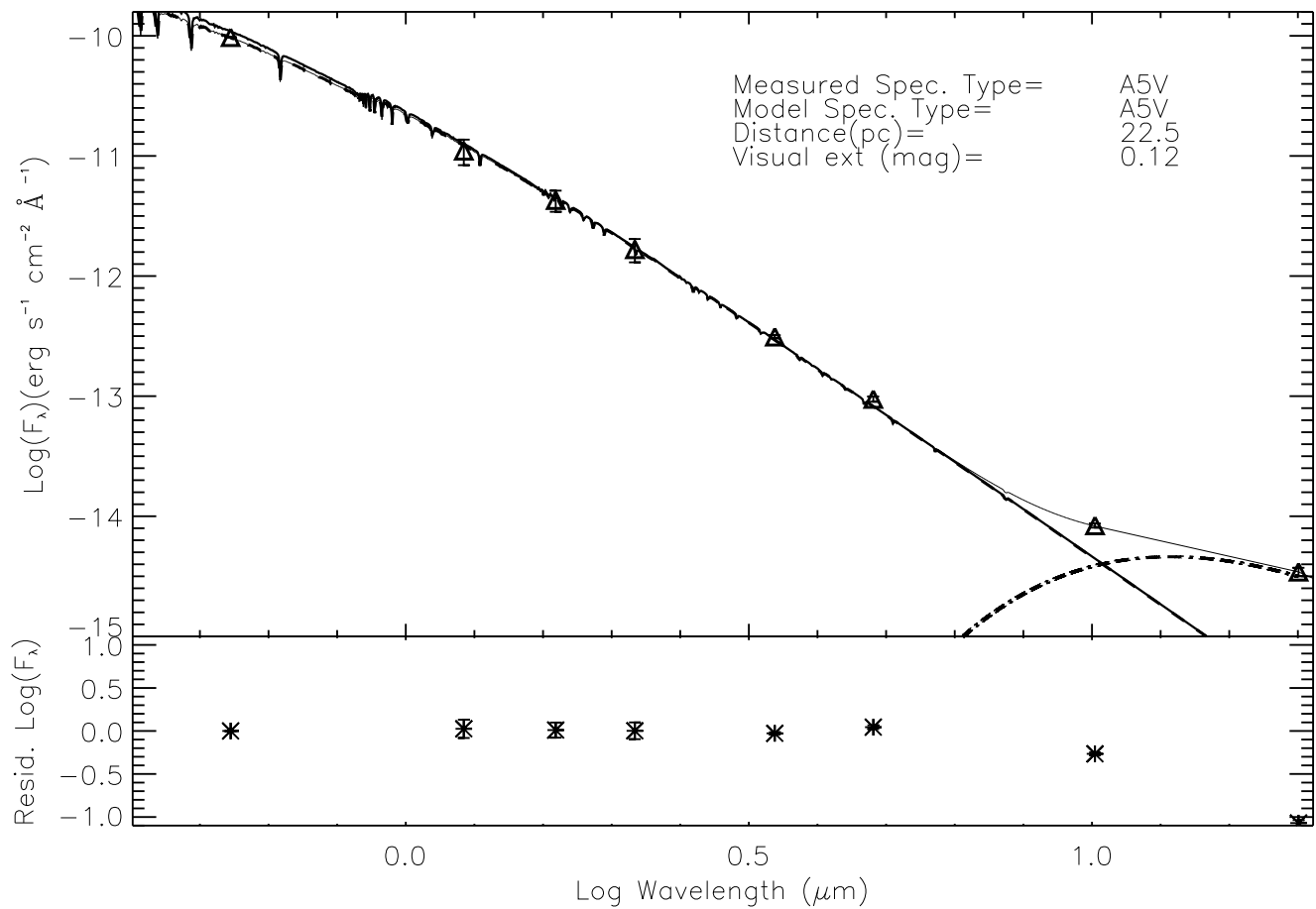}
    \caption{SED of $\beta$ Pictoris using GSC, $2$MASS and Backman
    \etal \ (1992) measurements.  Like most of our candidate sources,
    $\beta$ Pictoris shows an excess at longer mid-IR wavelengths.
    $\beta$ Pictoris did show a small excess at 4.80 $\mu$m in Backman
    \etal \ (1992). The thick solid line is the Kurucz model
    ($T_{eff}$=8200 K, Log g=+4.29), the thick dashed is the Kurucz model
    with extinction applied, the dot-dash line is the SED of the
    blackbody component with T=223 K, and the thin line is the
    combination of the Kurucz model with extinction plus the SED of
    the disk component. The residual plots below are shown before
    the addition of the blackbody component. They demonstrate that
    at longer wavelengths $\beta$ Pictoris deviates from the
    model photosphere.}
    \label{beta}
\end{figure}

An SED of $\beta$ Pictoris, the prototype debris disk system, is shown
in Figure~\ref{beta} for comparison to the excess
candidates. Figure~\ref{beta} also shows the appropriate Kurucz
model. The SED is comprised of L (3.45 $\mu$m), M (4.50 $\mu$m), N
(10.1 $\mu$m), and Q (20 $\mu$m) measurements from Backman \etal \
(1992).  The Q band measurement was included so that the number of
data points and the wavelength range was similar for both $\beta$
Pictoris's model and the G18 stars (8 measurements and 4 free
parameters in the model). The GSC and $2$MASS measurements were used
to model $\beta$ Pictoris in a manner equivalent to that for the G18
stars. Comparing the $2$MASS photometry to near-IR measurements in the
literature for $\beta$ Pictoris shows that the near-IR photometry is
consistent, but $2$MASS data have much larger uncertainties,
approximately 20$\%$. Our spectrophotometric distance of 22.5 $\pm$
3.0 pc for $\beta$ Pictoris is consistent with the Hipparcos distance
of 19.28 $\pm$ 0.19 pc.

In order to estimate the sizes and temperatures of any dust disks for
the G18 stars, we modeled the excess in each star as a single black
body with temperature and fractional bolometric disk-to-star
luminosity ratio, $\frac{L_{IR}}{L_{*}}$, as free parameters. This
assumption, although simplistic, yields a rough approximation to the
temperatures and $\frac{L_{IR}}{L_{*}}$ of the dust disks, given the
limited number of photometric points available. We searched a grid of
blackbody temperatures and fractional disk luminosities to find best
fits to the observed data. The temperature grid ranged from 100 K to
1100 K in steps of 1 K. This model was used to estimate the most
probable parameters for the dust disks. For our model of $\beta$
Pictoris, we found a single temperature for the excess to be 223$^{+
4}_{- 4}$ K with the disk emitting 0.0019$^{+ 0.0002}_{- 0.0002}$ the
luminosity of the star. In comparison, Gillett (1986) fit a single
temperature blackbody to both mid-IR $\textit{and}$ far-IR data, from
which they infer a temperature of 103 K and a fractional luminosity of
$0.003$ for $\beta$ Pictoris. By fitting the mid-IR data alone we may
be overestimating the disk temperature or just measuring the hotter
inner portion of a larger disk. Chen \& Jura (2001) found the color
temperature of $\zeta$ Leporis to be 370 K, but they note that if they
only fit the data at wavelengths greater than 10 $\mu$m the disk
temperature is better fit by a 230 K blackbody.  Backman \etal \
(1992) use a two disk component for $\beta$ Pictoris and find the
inner disk temperatures to range from 200-400+ K depending on the
model used. Therefore the temperature we estimate for the debris disk
is consistent with the range of disk temperatures given in Backman
\etal \ (1992) from a variety of more sophisticated models.

\begin{figure}
    \psfig{file=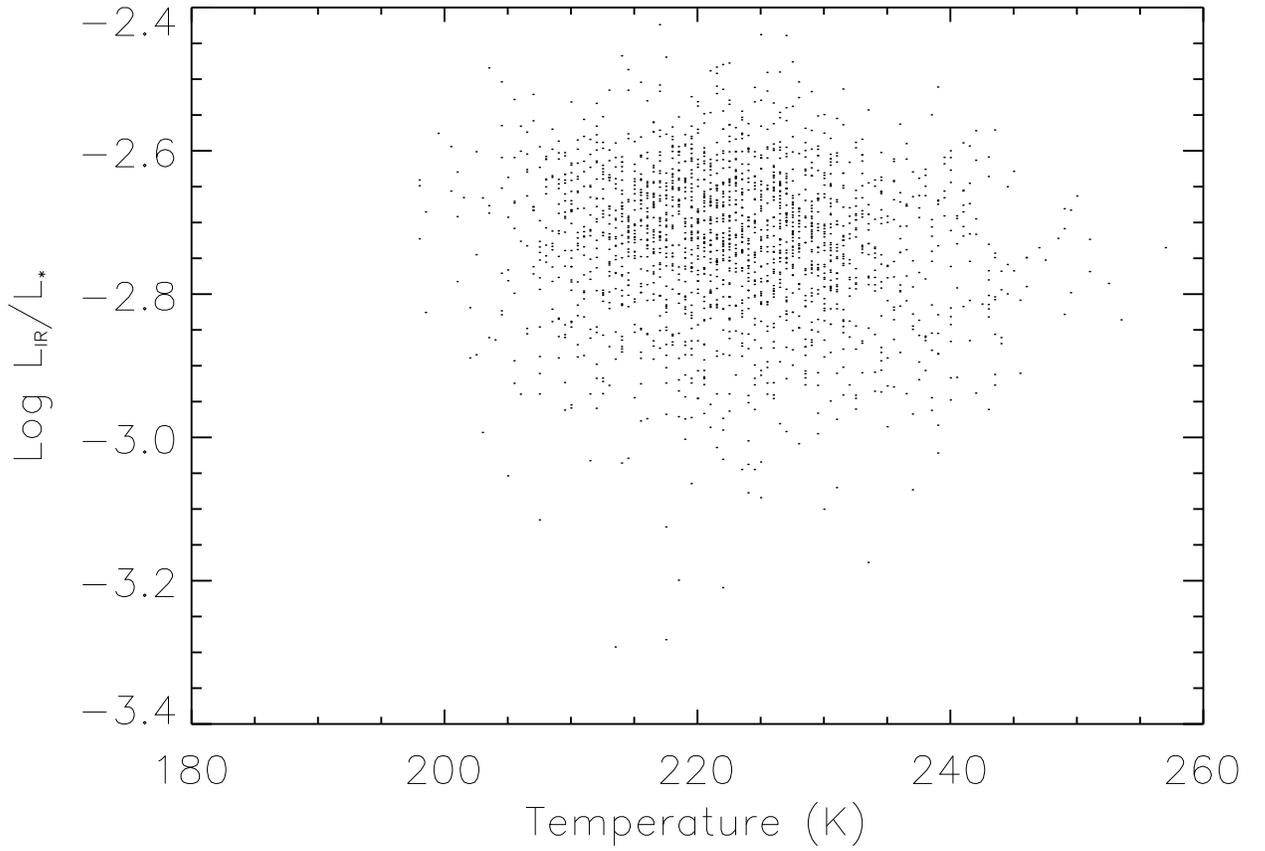}
    \caption{Plot of the Monte Carlo simulation for
     $\beta$ Pictoris.  The simulation demonstrates that temperature
     and fractional luminosity are well constrained
     to 223$^{+ 4}_{- 4}$ and fractional luminosity 0.0019$^{+
     0.0002}_{- 0.0002}$.}
    \label{cont1}
\end{figure}

\begin{figure}
    \psfig{file=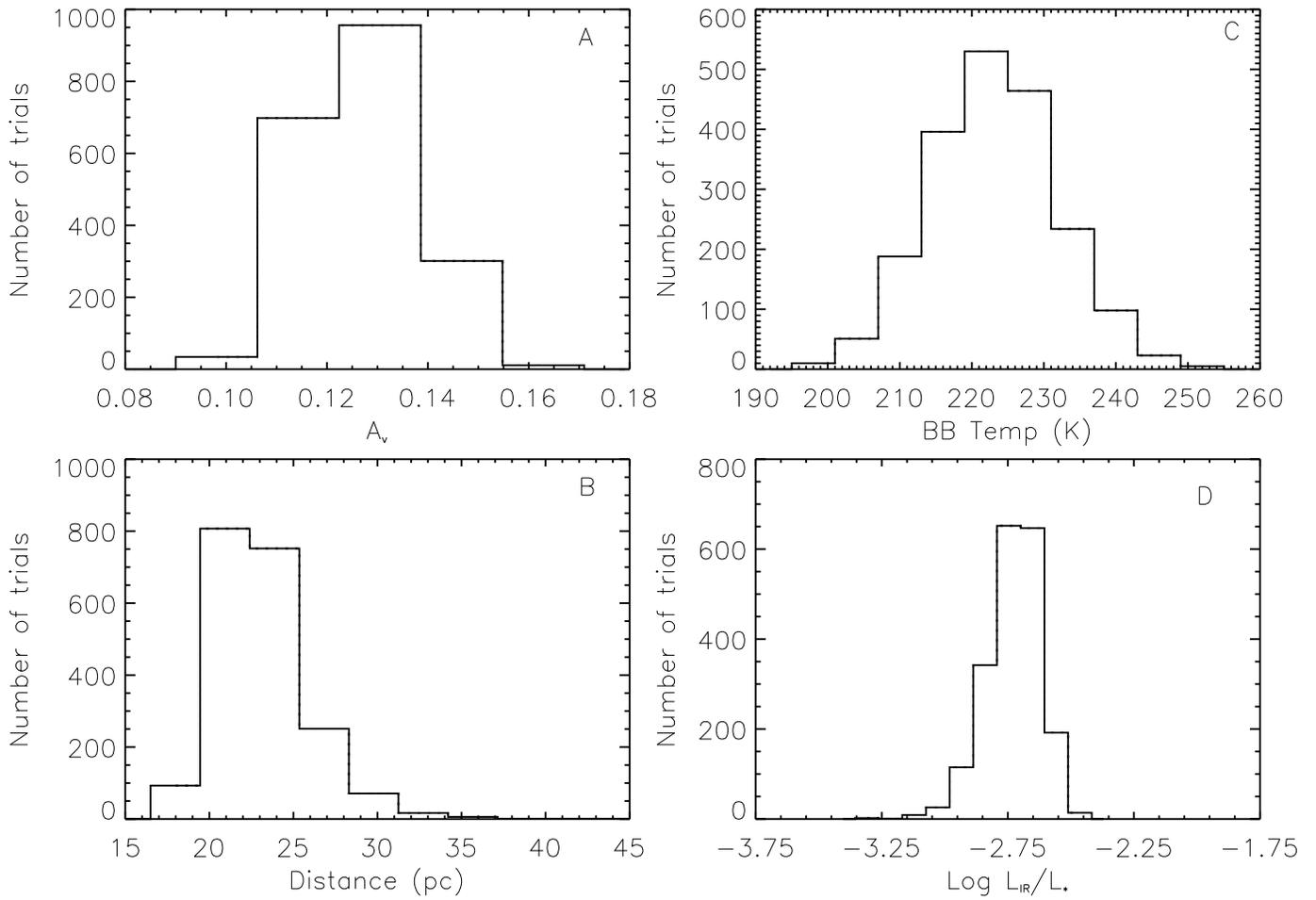}
    \caption{Histograms of (A) derived extinction,  (B) spectrophotometric
    distance, (C) blackbody temperature and (D) fractional luminosity for
    $\beta$ Pictoris using a Monte Carlo simulation.}
    \label{bhis}
\end{figure}

We investigated the likelihood of other disk temperature/fractional
luminosity combinations using a Monte Carlo simulation to determine
the uncertainties on our derived disk and star parameters. The Monte
Carlo code adds Gaussian noise to each of the photometric data points
(based on their 1 $\sigma$ photometric errors) and then re-computes
the most probable distance, visual extinction, blackbody temperature,
and fractional blackbody luminosity. Figure~\ref{cont1} shows a plot
of the Monte Carlo simulation results for $\beta$ Pic. The photometry
and model constrain the temperature and fractional luminosity to a
small region of the parameter space. Even a few near and mid-IR
photometric measurements effectively constrain the temperatures and
fractional disk luminosities. Figure~\ref{bhis} shows the distribution
functions of distance (before adding the 15\% stellar radii
uncertainty), visual extinction, blackbody temperature, and fractional
luminosity for $\beta$ Pictoris. Both the blackbody temperature and
fractional luminosity are well constrained in this
simulation. Spectrophotometric uncertainties are primarily due to the
assumption for the stellar radii uncertainties but are still well
constrained. With only the visual measurement in the optical, there is
some variation in extinction but for most stars the extinction is
low. Distributions for all the derived parameters are nearly Gaussian.

We then conducted the same analysis on our candidate stars. In a small
fraction of the Monte Carlo simulations, $\leq$ 5 $\%$ for most stars,
the best fit parameters require a fractional disk-to-star luminosity
ratio greater than unity. We discard these Monte Carlo iterations as
unphysical. In order for the luminosity of the disk to exceed the
luminosity of the star, the star would have to be embedded in a region
which required re-emission of absorbed stellar light, or generation of
energy by the disk. This would occur in early protostars, Class 0 or
I, but not in more evolved pre main-sequence stars. Three stars,
G284.1744-00.5141, G284.0110-00.1208, and G283.9773-00.3948, had
greater than 10$\%$ of their Monte Carlo simulations yield unphysical
fractional disk luminosities. These 3 stars have an excess only at 8
$\mu$m which provides weaker constraints on the fractional disk
luminosity and temperature. Table 3 lists derived parameters and their
uncertainties based on the median values of the Monte Carlo
simulation.

\begin{figure}
    \psfig{file=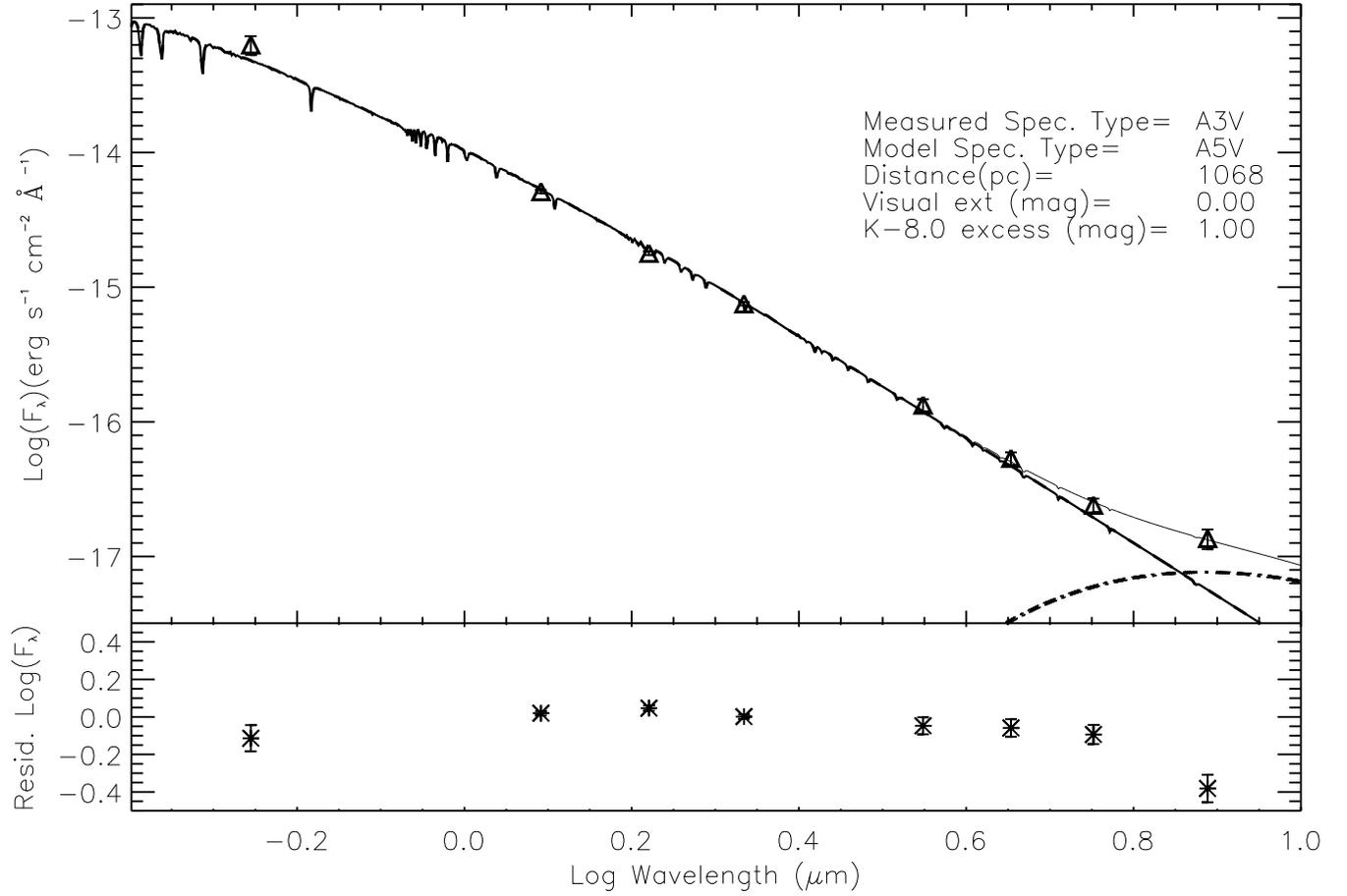}
    \caption{SED of G284.3535-00.2021, one of our candidate $\beta$
    Pictoris analogs. The thick solid line is the Kurucz model, the
    dot-dash line is the SED of the disk component, and the thin line
    is the combination of the Kurucz model with extinction plus the
    SED of the disk component. The residual plots below are shown
    before the addition of the blackbody component. G284.3535-00.2021
    shows a significant deviation from the photosphere at 8 $\mu$m.}
    \label{exce}
\end{figure}

Figure~\ref{exce} shows the SED for one of our G18 stars, the A3V star
G284.3535-00.2021. The thick solid line is the Kurucz model, the
dot-dash line is the SED of the disk component, and the thin line is
the combination of the Kurucz model with extinction plus the SED of
the disk component.

\begin{figure}
    \psfig{file=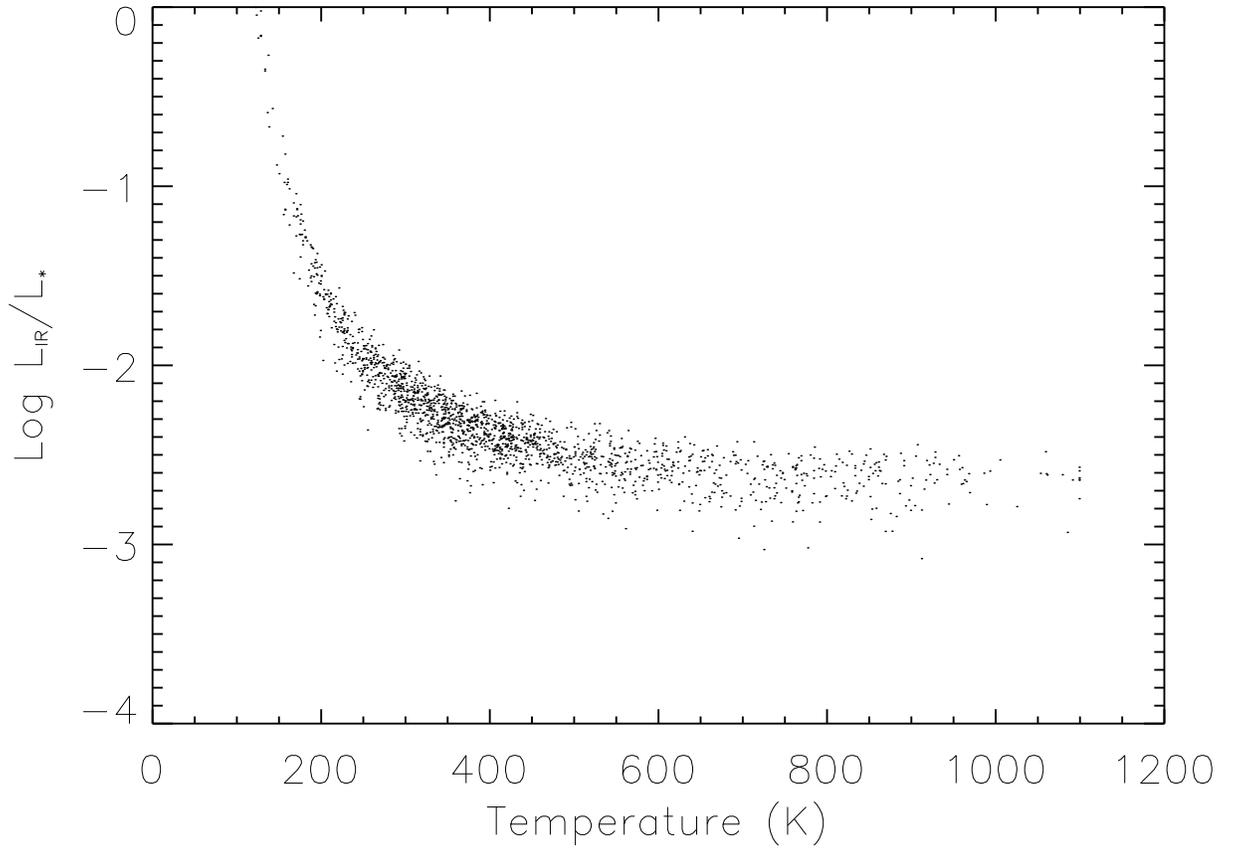}
    \caption{Scatter plot of the best fit blackbody temperature versus
     fractional disk-to-star luminosity ration based on 2000 Monte
     Carlo simulations for G284.3535-00.2021.  A wide range of
     temperatures and fractional luminosities are allowed for the given
     photometric measurements, but these allowed values are
     constrained to a narrow region of parameter space. The majority of
     the simulations indicate preferred disk temperatures of 315-440 K
     and fractional disk-to-star luminosities of 0.0033 to 0.0057 as
     show in the histograms in Figure~\ref{ghis}.}
    \label{cont2}
\end{figure}

Figure~\ref{cont2} shows the results for 2000 Monte Carlo
iterations, demonstrating that higher temperatures paired with
lower fractional luminosities occur more readily that lower
temperatures with higher fractional luminosities. This plot of
fractional disk luminosity versus temperature demonstrates that
there is an anti-correlation between fractional disk luminosity
and temperature. When blackbody temperature is high, the
fractional disk luminosity is low, and when temperature is low
fractional disk luminosity is high. The median temperature from
the simulations is 385 K (i.e., $\sim$$160$ K warmer than $\beta$
Pictoris), consistent with the observation that the excess appears
at shorter wavelengths in G284.3535-00.2021. The models constrain
the temperature to 385$^{+ 70}_{- 56}$ K and the fractional disk
luminosity to $0.0043$$^{+ 0.0014}_{- 0.0010}$.

\begin{figure}
    \psfig{file=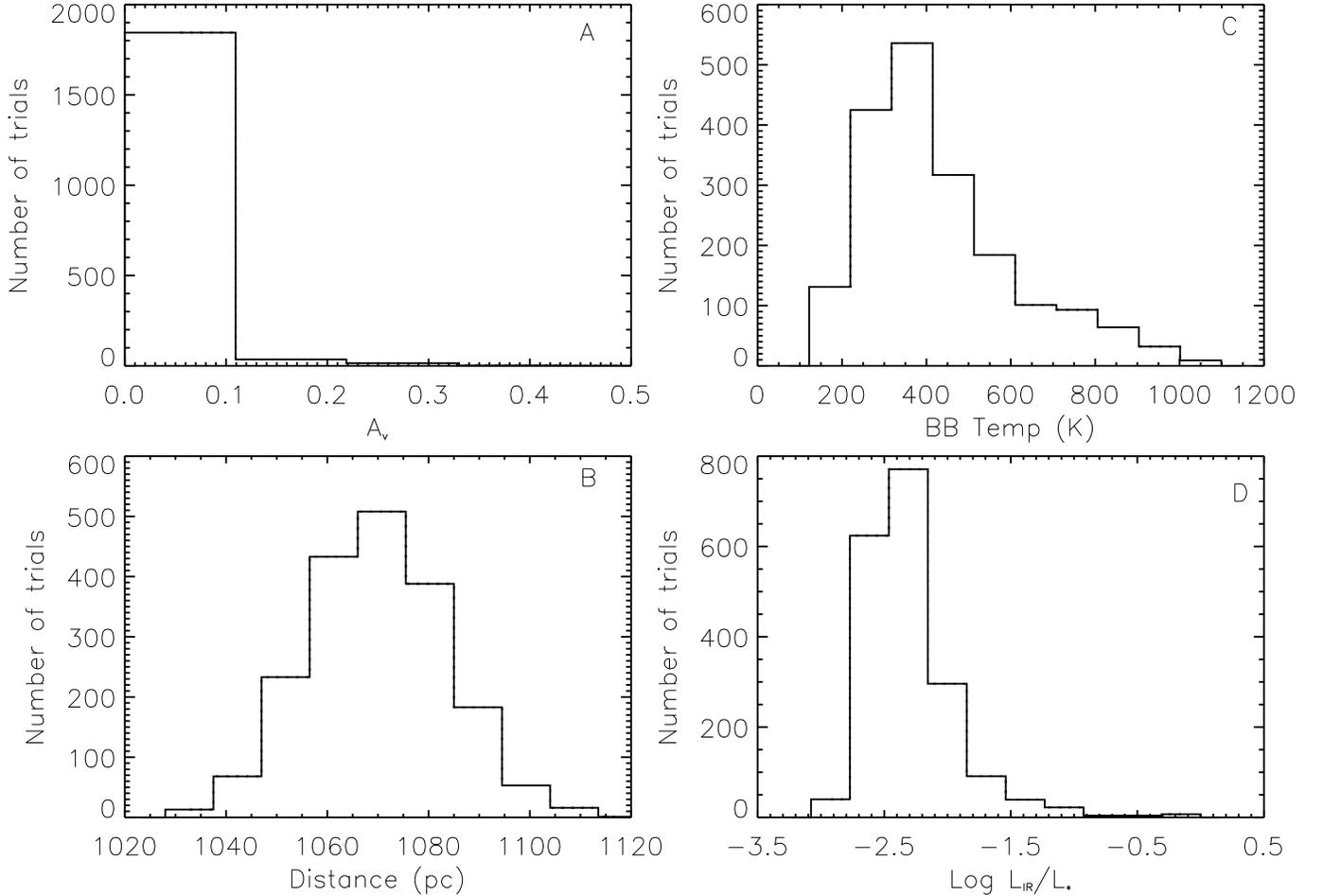}
    \caption{Histograms of (A) derived extinction, (B) spectrophotometric
    distance, (C) blackbody temperature and (D) fractional luminosity for
    G284.3535-00.2021.  The distributions for the derived parameters
    constrain the possible values, but the dispersion is larger than
    in the simulations of $\beta$ Pic because the models are not as
    tightly constrained without data longward of 8 $\mu$m.}
    \label{ghis}
\end{figure}

Figure~\ref{ghis} shows the distribution of distance (before
adding the 15\% stellar radii uncertainty), visual extinction,
temperature, and fractional luminosity for G284.3535-00.2021
determined from Monte Carlo simulations. This distribution is much
broader than that for $\beta$ Pictoris due to our limited ability
to constrain maximum possible temperatures with data shortward of
8 $\mu$m. Longer wavelength data, in both the mid and far-IR,
would allow us to better constrain the temperature and fractional
luminosity.

\begin{figure}
    \psfig{file=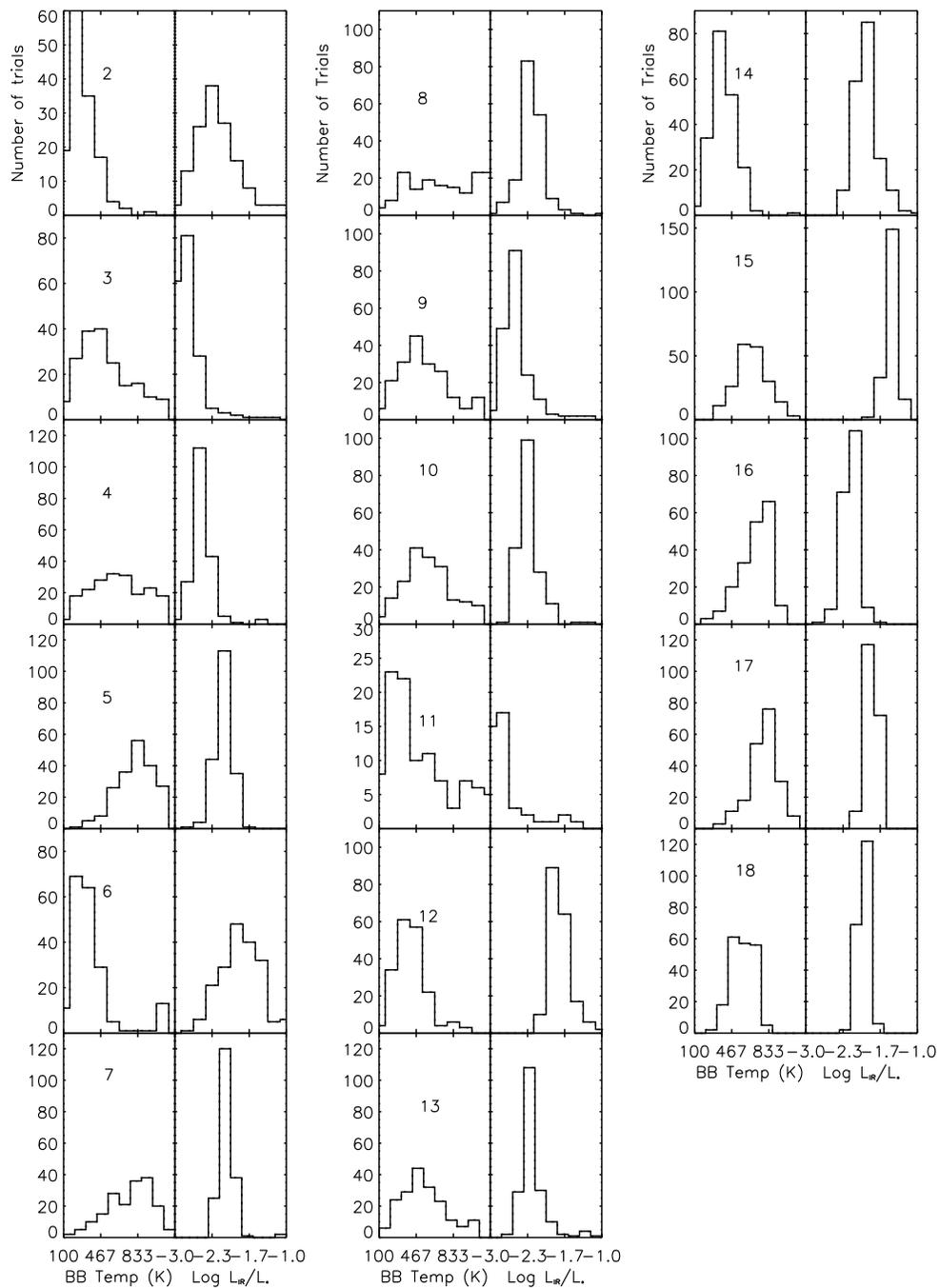,height=7.5in}
    \caption{Histograms of both temperature and fractional luminosity
    for the remaining 17 stars in Table 3. The reference number in
    the figure refers to the star ID in this paper.}
    \label{ahis}
\end{figure}

\begin{figure}
    \psfig{file=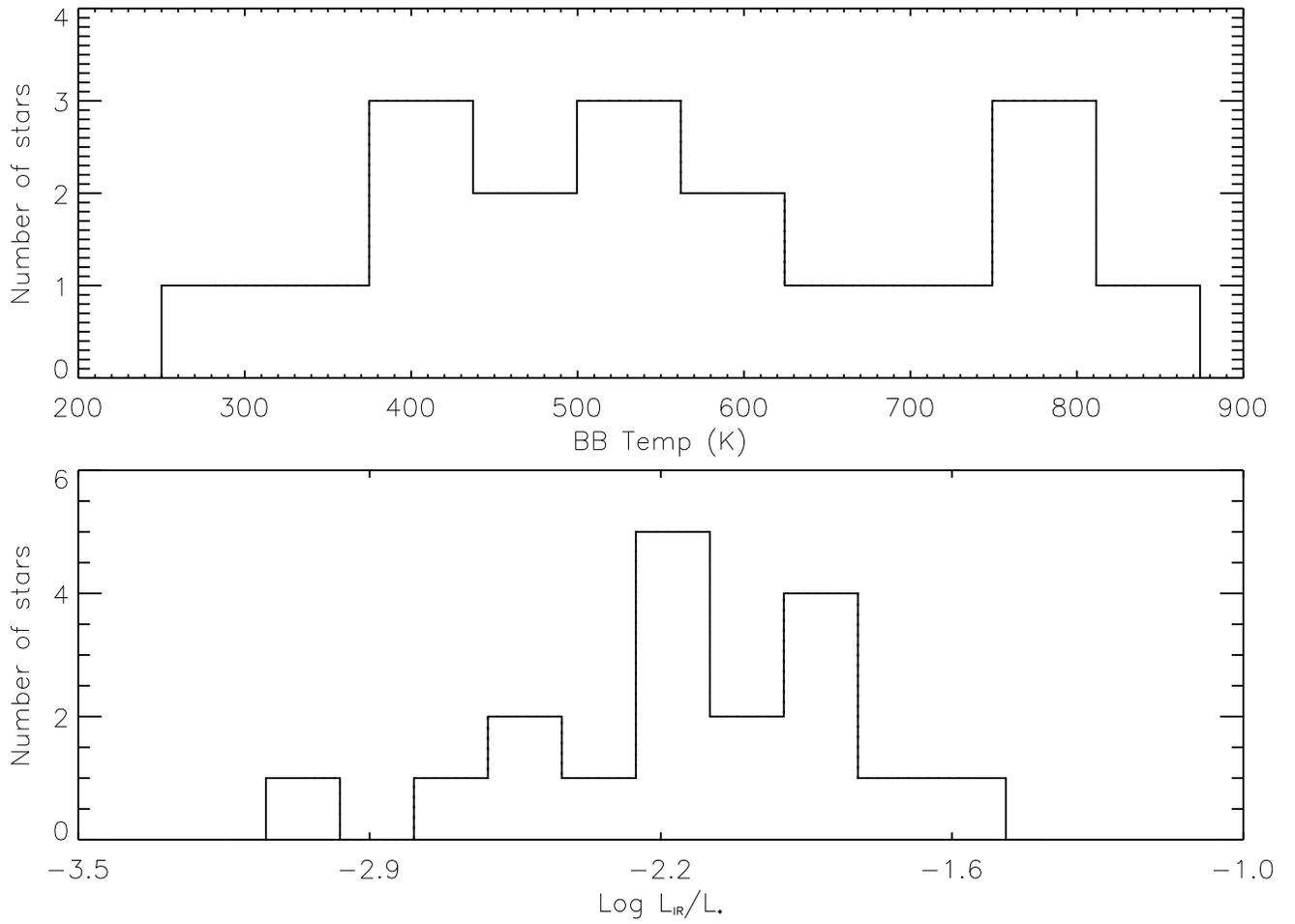}
    \caption{Histograms of both temperature and fractional luminosity
    for the median values derived for the G18 stars in Table 3. Although all the stars
    exhibit a mid-IR excess, there is a broad range of both temperatures and fractional luminosities for
    sample.}
    \label{dist}
\end{figure}

Table 3 lists the derived parameters and their uncertainties for all
18 main-sequence, non O and B stars with an excess (in the G18
sample).  The temperatures for the excesses range from 220-820 K with
fractional luminosities ranging from 10$^{-3}$ to 10$^{-2}$.
Figure~\ref{ahis} shows both temperature and fractional luminosity
ratio histograms for the remaining 17 stars in the G18 sample.  For
star 11 a large portion of the fractional luminosity ratio was below
10$^{-3}$ and is therefore off the scale. This is because there is a
significant excess at only [8.0]. G284.0110-00.1208 and
G284.0719-00.1637 have broad distributions, again due to the weaker
constraints on the excess. Figure~\ref{ahis} demonstrates the large
variation in temperature and fractional luminosity distribution
between stars.  Figure~\ref{dist} shows the distribution of both
temperature and fractional luminosity for all G18 stars. The median
temperature for the sample is $\sim500$ K. There does not appear to be
a correlation between either temperature or fractional disk luminosity
with spectral type but the sample is small. Future work may reveal a
correlation of either disk temperature or fractional luminosity with
spectral type.

\subsection{Protostellar Disk or Debris Disk?}

Lagrange \etal \ (2000) suggest a definition of debris disks using
four criteria. These criteria include the bolometric fractional
disk-to-star luminosity $\leq$ 10$^{-3}$, the mass of the gas and dust
to be below 10$^{-2}$ \mo , the dust mass significantly greater than
the gas mass, and the grain destruction time much less than the
stellar age. Our limited data only allow comparisons based on the
fractional bolometric luminosity.

Since $\frac{L_{IR}}{L_{*}}$ $\leq$ 0.1 for all of the G18 stars,
these stars are debris disks or may be in a transitionary phase from
protostellar to debris disk. Although these stars have blue spectra
which lack emission lines, except G284.3417-00.2049 which exhibits
H$\beta$ emission, and appear to be main-sequence, we can not rule out
H$\alpha$ emission. Weak H$\alpha$ emission could be present, without
H$\beta$ emission, which could imply that the star is pre
main-sequence. We can not rule out the possibility that the stars may
have primordial disks, but the disks are most likely to be debris
disks. Protostars typically have $\frac{L_{IR}}{L_{*}}$ $\geq$ 0.1 and
debris disks typically have $\frac{L_{IR}}{L_{*}}$ $\sim10^{-3}$
(Lagrange \etal \ 2000; Backman \& Paresce 1993). Since
G284.1744-00.5141, G284.0110-00.1208, and G283.9773-00.3948 had a
large number of their Monte Carlo iterations discarded because they
were unphysical, their fractional disk luminosities may be greater
than tabulated in Table 3. These stars may be protostellar in nature
since the models allow higher fractional luminosities, but their
derived luminosities, omitting the unphysical simulations, are still
consistent with debris disks. Mid-IR spectral analysis of the stars in
Table 3 would reveal whether the 10 $\mu$m silicate feature is present
and would be a useful tool in the characterization of mid-IR excesses
around main-sequence stars. The absorption (embedded protostar),
emission (pre main-sequence), or lack of the 10 $\mu$m silicate
feature (debris disk) is loosely related to the evolutionary
progression of a circumstellar disk (Kessler-Silacci \etal \
2005). However, the debris disks studied by Kessler-Silacci \etal \
(2005) contained debris disks that only exhibited far-IR
excesses. $\beta$ Pictoris, a mid-IR debris disk system, does have a
10 $\mu$m silicate emission feature. A useful comparison would be to
determine whether this sample of mid-IR debris disk systems also
exhibits a 10 $\mu$m silicate emission feature.

\section{G284.2642-00.3156}

\begin{figure}
    \psfig{file=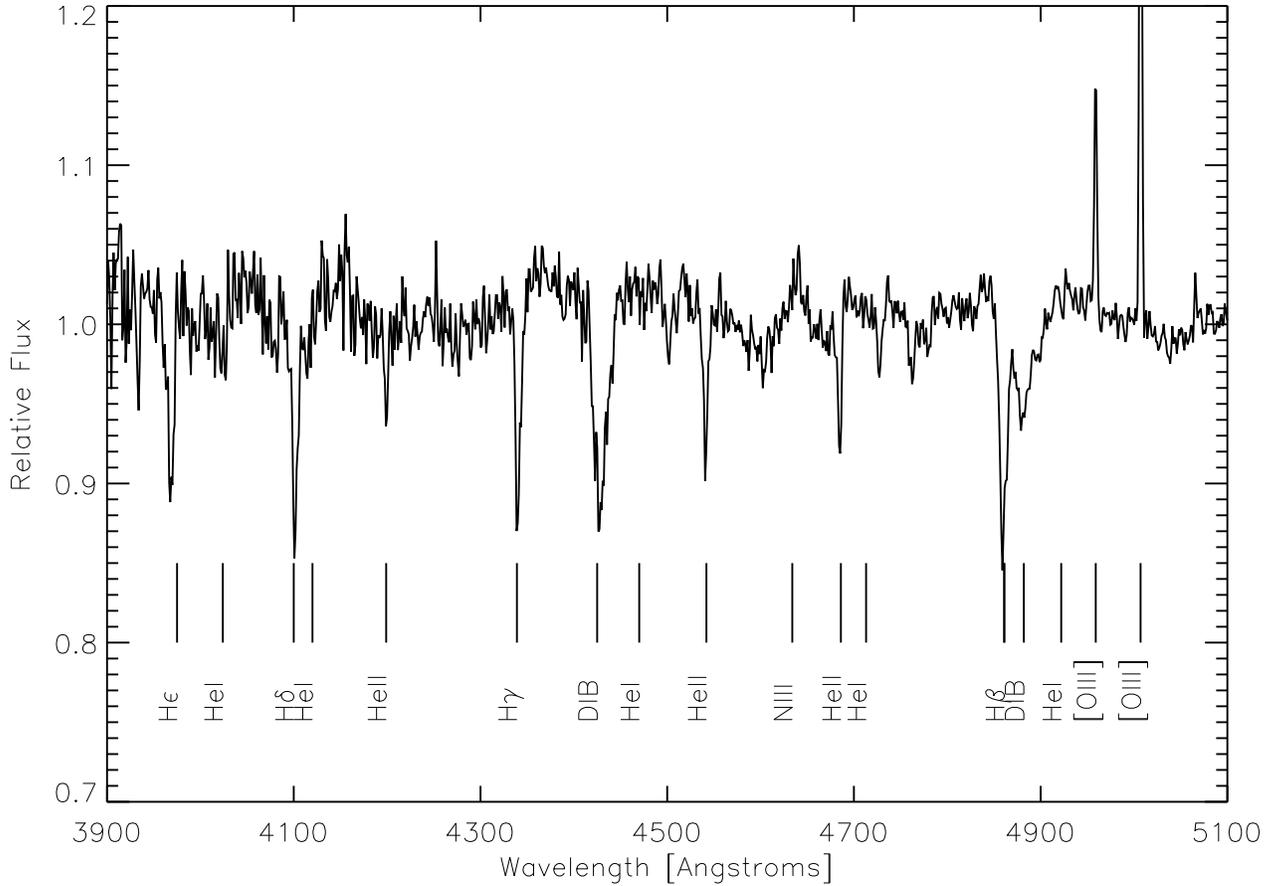}
    \caption{Optical Spectrum of G284.2642-00.3156 with stellar
    hydrogen and helium lines marked. The lack of strong \ion{He}{1}
    absorption lines but strong \ion{He}{2} lines make this a very
    early O-star, probably an O4. The diffuse interstellar band
    feature at 4430 \AA \ and 4882 \AA \ is consistent with the high
    extinction (A$_{V}$=5.6) to this star (Herbig 1995). The
    [\ion{O}{3}] emission lines reveal the presence of ionized gas
    near the star.}
    \label{spec}
\end{figure}

One especially interesting star that exhibited a mid-IR excess is
G284.2642-00.3156. The optical spectrum of this early type star is
shown in Figure~\ref{spec}. Using the OB star spectral atlas of
Walborn \& Fitzpatrick (1990) we classify this star as O4V((f)). The
classification is based on the relative strength of \ion{He}{1},
\ion{He}{2}, and the hydrogen Balmer series. Figure~\ref{spec} shows
that for this star weak \ion{N}{3} 4634-40-42 \AA \ appears in
emission and a strong \ion{He}{2} 4686 \AA \ appears in absorption,
which implies that the star is earlier than O5. \ion{He}{1} is present
in absorption but is very weak, which implies that the star is later
than O3 since an O3 lacks \ion{He}{1} absorption. We did not see any
\ion{N}{4} 4058 \AA \ emission or any significant \ion{N}{5} 4604 \AA
\ or 4620 \AA \ absorption, and \ion{He}{2} 4686 \AA \ is in
absorption. These imply that the star is not a supergiant or
giant.

\begin{figure}
    \psfig{file=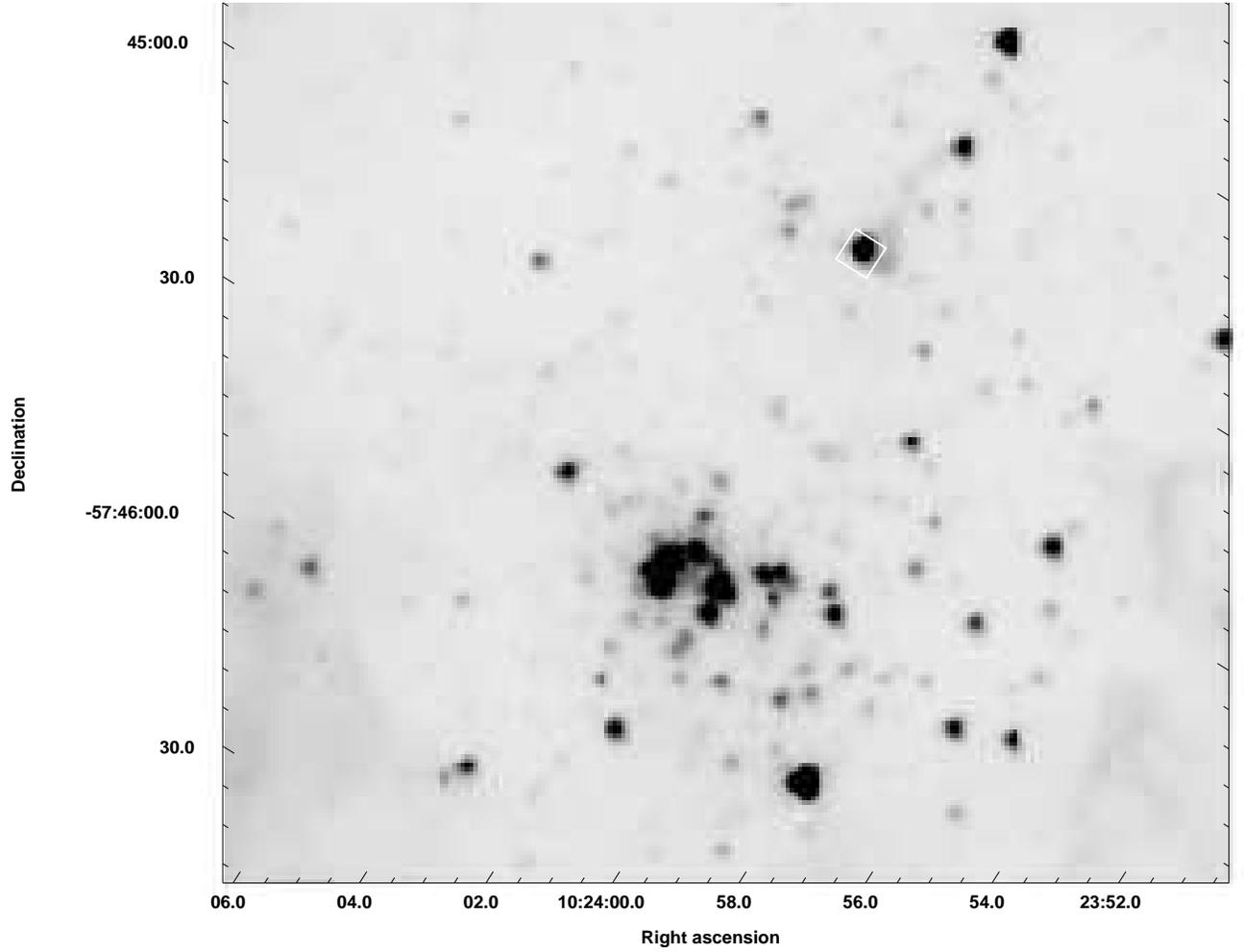,width=8.5in,bbllx=130bp,bblly=220bp,bburx=500bp,bbury=500bp,clip=.}
    \caption{[3.6] \textit{Spitzer} IRAC image showing both Westerlund
  2 and the O4V((f)) star. The position of the O star is shown by a
  white box. Diffuse emission in this band is attributed to known PAH
  features.}
  \label{ostar1}
\end{figure}

\begin{figure}
    \psfig{file=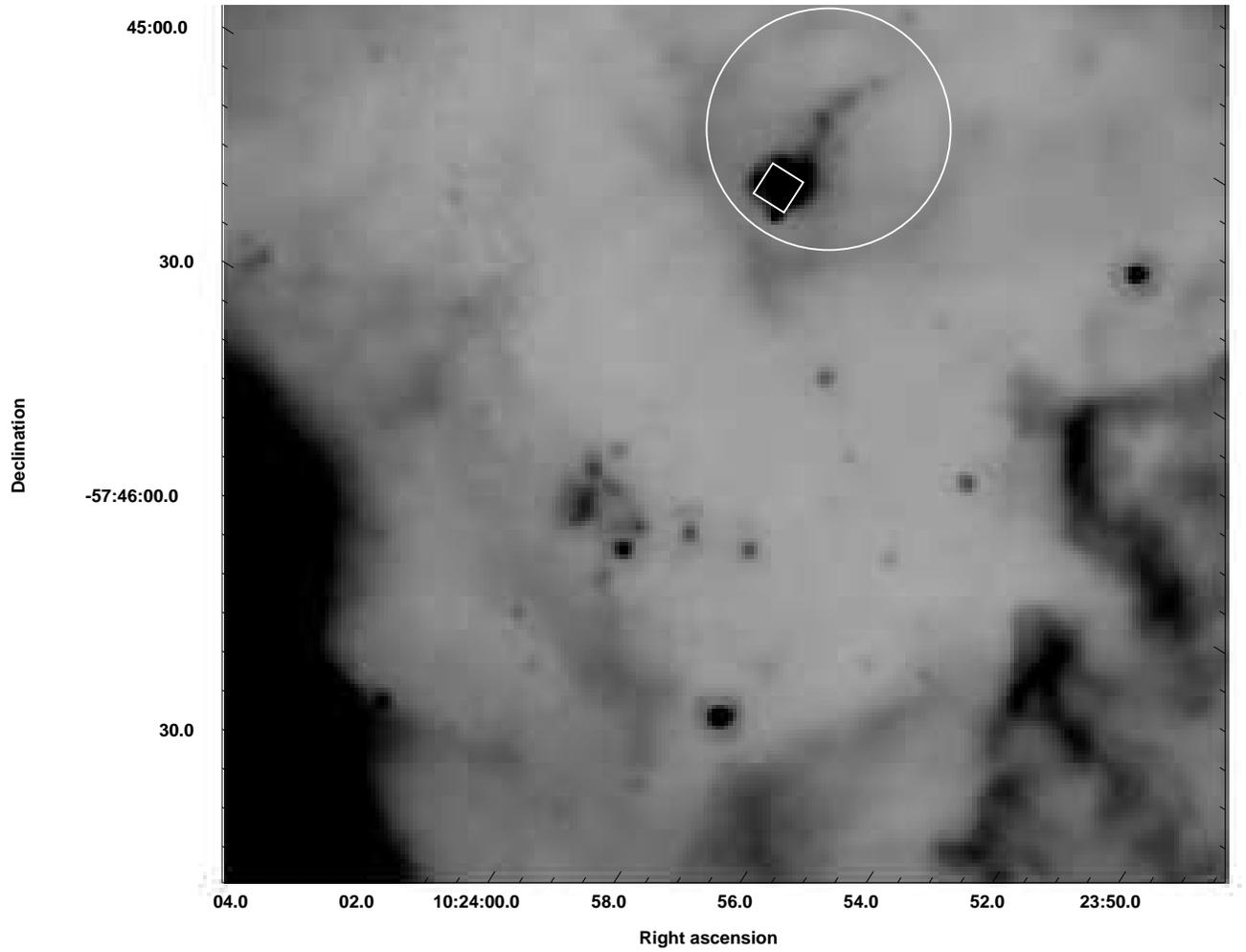,width=8.5in,bbllx=130bp,bblly=220bp,bburx=500bp,bbury=500bp,clip=.}
  \caption[Ostar] {[8.0] \textit{Spitzer} IRAC image showing both
  Westerlund 2 and the O4V((f)) star. The position of the O-star is
  shown by a white box. The [8.0] image shows a circular emission ring
  around the star, although the star is not centered within this
  ring. The ring and the irregular linear feature extending to the NW
  from the star may trace molecular material or dust illuminated by
  the star. The diffuse emission is attributed to known PAH features
  in the [8.0] band.}
  \label{ostar4}
\end{figure}

Figure~\ref{ostar1} and Figure~\ref{ostar4} show GLIMPSE [3.6] and
[8.0] images in the region surrounding G284.2642-00.3156, including
Westerlund 2. G284.2642-00.3156 is marked by the white box. The [3.6]
image shows that G284.2642-00.3156 is surrounded by a grouping of
faint stars which are partially blended with the O4 star at the 1.22
arcsecond angular resolution of the IRAC 3.6 $\mu$m array.  We
determine a spectrophotometric distance to this star of 3233$^{+
540}_{- 535}$ pc with a visual extinction of 5.63$^{+ 0.01}_{-
0.30}$. The [8.0] image shows a nearly circular ring with radius of
$\sim$$38$$^{\prime \prime}$ surrounding the O star. At a distance of
3.5 kpc, the ring around the star would be 0.65 pc in diameter. This
could be a wind blown shell illuminated by the star. There is also an
irregular linear feature extending 29$^{\prime \prime}$ from the star
in the northwest direction. This feature may be a nearby cloud
illuminated by the star, or may trace the relative motion of the star
through the inter-stellar medium surrounding Westerlund 2.

Using the images of Moffat \etal \ (1991) we identify
G284.2642-00.3156 as their star $\#$$18$, which they find to be the
most luminous star in the cluster.  They type this star as an O7V, but
they note that its absolute luminosity is brighter than that of an
O7V. Moffat \etal \ conclude that since the star appears isolated, it
must, by color and luminosity arguments, be a supergiant if it lies at
the distance of the cluster. Our new classification results in an
earlier spectral type with an extinction which is still consistent
with cluster membership, but a distance that is not consistent with
the 7.9 kpc by Moffat \etal (1991). We note that the distance we find
for the O4V((f)) star is inconsistent with the Carraro \& Munari
(2004) cluster distance of 6.3 kpc but still consistent with the wide
variation of distances found for this cluster (Churchwell \etal \
2004).  If this star is a cluster member it would be the earliest
known member. Moffat \etal \ (1991) assumed a cluster age of 2-3 Myr
based on the presence of the O7 stars. The proximity of WR20b to the
cluster may imply an older age of 3-5 Myr if it is also a cluster
member (Shara \etal \ 1991). The lifetime for an O4 is on the order of
3 Myr (for a 60 \mo star with Z=0.02;Meynet \etal \
1994). G284.2642-00.3156 lies 1 arcminute from the cluster center,
which corresponds to a minimum projected separation of 1 pc if the
star is at a distance of 3.5 kpc. In order to travel 1 pc in 3 Myr or
less, a modest space velocity of $\geq$ 0.3 km s$^{-1}$ is
required. Thus the reddening, and projected separation of the O4 star
are consistent with cluster membership and origin. This implies that
Westerlund 2 may be younger and closer than previously
thought. However, G284.2642-00.3156 may have been formed subsequent to
the formation of Westerlund 2 as a result of triggered star formation
in surrounding molecular clouds (Deharveng \etal \ 2005, Oey \etal \
2005). Further studies will determine if this star is a cluster member
or a result of triggered star formation as well as the distance to
this cluster. High resolution spectra of both the O4 star and
Westerlund 2 would allow a measurement if their relative radial
velocities and would help establish whether this star is a cluster
member.

\section{Conclusion}

We found 33 stars which exhibited a mid-IR excess above photospheric
levels in the field surrounding RCW$49$. We combined our new spectral
classifications with known literature classifications and optical,
near, and mid-infrared photometry to model the spectral energy
distributions of 184 stars. Stars with mid-IR excesses span all
spectral classes from B to early K. We modeled the excess for each
star as a single component blackbody and found that for the G18 stars,
the excess is consistent with a debris disk or some transitionary
phase between primordial circumstellar disk and debris disk. For these
$18$ stars the additional black body component was found to have
fractional bolometric disk-to-star luminosity ratios,
$\frac{L_{IR}}{L_{*}}$, ranging from (10$^{-3}$ to 10$^{-2}$) with
temperatures ranging from 220 to 820 K. These temperatures and
fractional disk-to-star luminosities are consistent with warm inner
dust of debris disks which could be analogous to asteroid belt type
objects.

\acknowledgments

We would like to thank Michael Meyer for his comments and
suggestions. We would like to thank our anonymous referee for their
useful comments and suggestions. Support for this work, part of the
Spitzer Space Telescope Legacy Science Program, was provided by NASA
through Contract Numbers (institutions) 1224653 (UW), 1225025 (BU),
1224681 (UMd), 1224988 (SSI), 1242593 (UCB), 1253153 (UMn), 1253604
(UWy), 1256801 (UWW) by the Jet Propulsion Laboratory, California
Institute of Technology under NASA contract 1407. Brian Uzpen
acknowledges support by the Wyoming NASA Space Grant Consortium, NASA
Grant NGT-40102 40102, Wyoming NASA EPSCoR Grant NCC5-578 and 1253604.

\clearpage

\begin{deluxetable}{cccccccccccccccccccc}
\rotate
\tabletypesize{\tiny}
\setlength{\tabcolsep}{0.01in}
\tablewidth{8.0in}
\tablecaption{Photometric Parameters for Stars with Mid-IR Excesses}
\tablehead{
\colhead{\ } &
\colhead{ID} &
\colhead{2MASS ID} &
\colhead{K} &
\colhead{$\sigma$K} &
\colhead{V-K} &
\colhead{$\sigma$V-K} &
\colhead{J-K} &
\colhead{$\sigma$J-K} &
\colhead{H-K} &
\colhead{$\sigma$H-K} &
\colhead{K-[3.6]} &
\colhead{$\sigma$K-[3.6]} &
\colhead{K-[4.5]} &
\colhead{$\sigma$K-[4.5]} &
\colhead{K-[5.8]} &
\colhead{$\sigma$K-[5.8]} &
\colhead{K-[8.0]} &
\colhead{$\sigma$K-[8.0]} &
\colhead{Spec.} \\
\colhead{\ } &
\colhead{} &
\colhead{} &
\colhead{[mag]} &
\colhead{[mag]} &
\colhead{[mag]} &
\colhead{[mag]} &
\colhead{[mag]} &
\colhead{[mag]} &
\colhead{[mag]} &
\colhead{[mag]} &
\colhead{[mag]} &
\colhead{[mag]} &
\colhead{[mag]} &
\colhead{[mag]} &
\colhead{[mag]} &
\colhead{[mag]} &
\colhead{[mag]} &
\colhead{[mag]} &
\colhead{} \\
\colhead{(1)} &
\colhead{(2)} &
\colhead{(3)} &
\colhead{(4)} &
\colhead{(5)} &
\colhead{(6)} &
\colhead{(7)} &
\colhead{(8)} &
\colhead{(9)} &
\colhead{(10)} &
\colhead{(11)} &
\colhead{(12)} &
\colhead{(13)} &
\colhead{(14)} &
\colhead{(15)} &
\colhead{(16)} &
\colhead{(17)} &
\colhead{(18)} &
\colhead{(19)} &
\colhead{(20)}  }
\startdata
 1 & G284.3535--00.2021 & 10250358-5741409 & 11.89 & 0.03 & -0.05 & 0.17 &  0.07 & 0.04 &  0.10 & 0.04 &  0.14 & 0.04 &  0.15 & 0.05 &  0.24 & 0.07 &  0.97 & 0.14 & A3V \\
 2 & G284.1744--00.5141 & 10224039-5751461 & 12.47 & 0.03 & -0.00 & 0.43 &  0.06 & 0.04 &  0.05 & 0.04 &  0.02 & 0.04 &  0.01 & 0.05 &  0.13 & 0.15 &  0.89 & 0.12 & A2V \\
 3 & G283.8842--00.3361 & 10213378-5733229 & 11.97 & 0.02 & -0.13 & 0.17 &  0.02 & 0.03 &  0.00 & 0.03 &  0.18 & 0.03 &  0.12 & 0.04 &  0.22 & 0.07 &  0.73 & 0.09 & B9V \\
 4 & G284.1241--00.2429 & 10232673-5736249 & 10.92 & 0.02 &  0.86 & 0.16 &  0.11 & 0.03 &  0.03 & 0.03 &  0.14 & 0.03 &  0.14 & 0.04 &  0.22 & 0.05 &  0.53 & 0.06 & A7V \\
 5 & G284.0185--00.1803 & 10230188-5729511 & 12.66 & 0.03 &  1.98 & 0.44 &  0.55 & 0.04 &  0.11 & 0.04 &  0.14 & 0.04 &  0.08 & 0.05 &  0.45 & 0.11 &  0.45 & 0.08 & K0V \\
 6 & G284.0547--00.5695 & 10214145-5750415 & 12.73 & 0.04 &  0.99 & 0.43 &  0.21 & 0.05 &  0.07 & 0.05 &  0.10 & 0.05 &  0.11 & 0.06 &  0.40 & 0.09 &  1.27 & 0.23 & B8V \\
 7 & G284.0719--00.1637 & 10232601-5730435 & 12.60 & 0.03 &  1.66 & 0.43 &  0.43 & 0.04 &  0.13 & 0.04 &  0.22 & 0.04 &  0.26 & 0.05 &  0.28 & 0.09 &  0.65 & 0.08 & F3IV \\
 8 & G284.0110--00.1208 & 10231333-5726356 & 12.46 & 0.04 &  1.54 & 0.43 &  0.33 & 0.05 &  0.06 & 0.05 &  0.20 & 0.05 &  0.14 & 0.05 &  0.18 & 0.09 &  0.49 & 0.07 & F5IV \\
 9 & G284.2320--00.1670 & 10242577-5736016 & 11.24 & 0.02 &  0.61 & 0.43 &  0.15 & 0.03 &  0.06 & 0.03 &  0.09 & 0.03 &  0.11 & 0.03 &  0.25 & 0.05 &  0.61 & 0.04 & A5V  \\
10 & G284.0658--00.3254 & 10224485-5738431 & 11.51 & 0.02 &  1.35 & 0.43 &  0.34 & 0.03 &  0.07 & 0.03 &  0.11 & 0.03 &  0.08 & 0.04 &  0.22 & 0.06 &  0.55 & 0.06 & G0V  \\
11 & G283.9773--00.3948 & 10215463-5739217 & 12.38 & 0.02 &  0.02 & 0.23 &  0.04 & 0.03 &  0.00 & 0.04 &  0.03 & 0.04 &  0.05 & 0.05 & -0.01 & 0.08 &  0.37 & 0.08 & B8V  \\
12 & G283.9935--00.1944 & 10224911-5729455 & 12.21 & 0.02 &  1.68 & 0.43 &  0.45 & 0.03 &  0.11 & 0.03 &  0.19 & 0.04 &  0.13 & 0.05 &  0.30 & 0.07 &  1.05 & 0.06 & G5V  \\
13 & G283.9239--00.5103 & 10210640-5743271 & 12.09 & 0.02 &  0.97 & 0.43 &  0.27 & 0.03 &  0.03 & 0.04 &  0.12 & 0.04 &  0.14 & 0.04 &  0.22 & 0.07 &  0.70 & 0.08 & F5IV/V \\
14 & G283.9153--00.4337 & 10212181-5739183 & 12.80 & 0.03 &  1.34 & 0.43 &  0.30 & 0.04 &  0.07 & 0.04 &  0.05 & 0.05 &  0.07 & 0.06 &  0.41 & 0.12 &  0.94 & 0.13 & F3V  \\
15 & G284.0478--00.1686 & 10231576-5730119 & 13.33 & 0.05 &  \nodata & \nodata &  0.81 & 0.06 &  0.19 & 0.06 &  0.24 & 0.06 &  0.18 & 0.08 &  0.79 & 0.19 &  0.99 & 0.15 & K5V \\
16 & G283.9076--00.1997 & 10221550-5727151 & 10.90 & 0.02 &  \nodata & \nodata &  0.95 & 0.04 &  0.22 & 0.04 &  0.18 & 0.04 &  0.14 & 0.04 &  0.22 & 0.05 &  0.51 & 0.05 & G8IV \\
17 & G283.9040--00.3687 & 10213333-5735398 & 12.06 & 0.02 &  \nodata & \nodata &  0.83 & 0.03 &  0.17 & 0.03 &  0.18 & 0.04 &  0.14 & 0.05 &  0.41 & 0.07 &  0.58 & 0.08 & K5V  \\
18 & G284.3417--00.2049 & 10245840-5741272 & 11.57 & 0.04 &  \nodata & \nodata &  1.39 & 0.06 &  0.36 & 0.06 &  0.34 & 0.05 &  0.33 & 0.05 &  0.43 & 0.07 &  1.05 & 0.12 & F8Ve \\
\enddata
\tablerefs{ (1) Reference ID \# for this paper; (2) GLIMPSE
Catalog ID which is in Galactic coordinates; (3) 2MASS Catalog ID;
(4) 2MASS K magnitude; (5) $1\sigma$ uncertainty for K magnitude;
(6) Guide Star Catalog 2.2 V - 2MASS K; (7) $1\sigma$ uncertainty
for V-K; (8) 2MASS J - 2MASS K; (9) $1\sigma$ uncertainty for
J-K; (10) 2MASS H - 2MASS K; (11) $1\sigma$ uncertainty for H-K;
(12) 2MASS K - IRAC [3.6] using zero-points from (Cohen \etal \
2003) used in all colors; (13) $1\sigma$ uncertainty for K-[3.6];
(14) 2MASS K - IRAC [4.5]; (15) $1\sigma$ uncertainty for
K-[4.5]; (16) 2MASS K - IRAC [5.8]; (17) $1\sigma$ uncertainty
for K-[5.8]; (18) 2MASS K - IRAC [8.0]; (19) $1\sigma$
uncertainty for K-[8.0]; (20) Spectral type}
\end{deluxetable}

\begin{deluxetable}{ccccccccccccccccccccc}
\rotate
\tabletypesize{\tiny}
\setlength{\tabcolsep}{0.01in}
\tablewidth{8.5in}
\tablecaption{Other Stars with Mid-IR Excesses}
\tablehead{
\colhead{\ }&
\colhead{ID} &
\colhead{2MASS ID} &
\colhead{K} &
\colhead{$\sigma$K} &
\colhead{V-K} &
\colhead{$\sigma$V-K} &
\colhead{J-K} &
\colhead{$\sigma$J-K} &
\colhead{H-K} &
\colhead{$\sigma$H-K} &
\colhead{K-[3.6]} &
\colhead{$\sigma$K-[3.6]} &
\colhead{K-[4.5]} &
\colhead{$\sigma$K-[4.5]} &
\colhead{K-[5.8]} &
\colhead{$\sigma$K-[5.8]} &
\colhead{K-[8.0]} &
\colhead{$\sigma$K-[8.0]} &
\colhead{Spec.} &
\colhead{E(K-8)}\\
\colhead{\ } &
\colhead{} &
\colhead{} &
\colhead{[mag]} &
\colhead{[mag]} &
\colhead{[mag]} &
\colhead{[mag]} &
\colhead{[mag]} &
\colhead{[mag]} &
\colhead{[mag]} &
\colhead{[mag]} &
\colhead{[mag]} &
\colhead{[mag]} &
\colhead{[mag]} &
\colhead{[mag]} &
\colhead{[mag]} &
\colhead{[mag]} &
\colhead{[mag]} &
\colhead{[mag]} &
\colhead{} &
\colhead{[mag]} \\
\colhead{(1)} &
\colhead{(2)} &
\colhead{(3)} &
\colhead{(4)} &
\colhead{(5)} &
\colhead{(6)} &
\colhead{(7)} &
\colhead{(8)} &
\colhead{(9)} &
\colhead{(10)} &
\colhead{(11)} &
\colhead{(12)} &
\colhead{(13)} &
\colhead{(14)} &
\colhead{(15)} &
\colhead{(16)} &
\colhead{(17)} &
\colhead{(18)} &
\colhead{(19)} &
\colhead{(20)} &
\colhead{(21)}  }
\startdata
19 & G283.9403--00.2636 & 10221243-5731324 &  9.81 & 0.02 &  0.52 & 0.05 &  0.07 & 0.03 &  0.06 & 0.03 &  0.51 & 0.03 &  0.71 & 0.03 &  0.92 & 0.03 &  1.32 & 0.02 & B2V(weak Be) & 1.43 \\
20 & G284.1728--00.2039 & 10235451-5736004 & 10.63 & 0.02 &  0.49 & 0.06 &  0.08 & 0.03 &  0.00 & 0.03 &  0.54 & 0.03 &  0.61 & 0.03 &  0.73 & 0.05 &  1.03 & 0.09 & B1V (weak Be) & 1.08 \\
21 & G284.1277--00.5835 & 10220574-5753460 & 10.60 & 0.02 &  3.91 & 0.44 &  0.77 & 0.03 &  0.28 & 0.03 &  0.27 & 0.03 &  0.41 & 0.03 &  0.49 & 0.04 &  1.48 & 0.06 & B5Vneb      & 1.23 \\
22 & G284.0335--00.2091 & 10230066-5731474 & 12.74 & 0.03 &  1.68 & 0.44 &  0.39 & 0.04 &  0.11 & 0.04 &  0.15 & 0.04 &  0.22 & 0.05 &  0.29 & 0.12 &  0.60 & 0.12 & F5II        & 0.57 \\
23 & G284.2642--00.3156 & 10240243-5744359 &  8.65 & 0.03 &  \nodata & \nodata &  0.84 & 0.04 &  0.16 & 0.05 &  0.40 & 0.05 &  0.75 & 0.05 &  1.50 & 0.03 &  3.84 & 0.03 & O4V((f))         & 4.08 \\
24 & G284.0501--00.2464 & 10225795-5734130 & 11.50 & 0.02 &  \nodata & \nodata &  0.98 & 0.03 &  0.23 & 0.03 &  0.23 & 0.03 &  0.19 & 0.03 &  0.29 & 0.06 &  0.75 & 0.06 & G8II        & 0.70 \\
25 & G283.9567+00.1258 & 10235162-5712213 & 13.04 & 0.03 &  \nodata & \nodata &  0.38 & 0.04 &  0.08 & 0.04 &  0.18 & 0.05 &  0.11 & 0.06 &  0.43 & 0.17 &  0.59 & 0.09 & G8III       & 0.54 \\
26 & G283.9776+00.1738 & 10241078-5710353 & 13.03 & 0.02 &  \nodata & \nodata &  0.39 & 0.03 &  0.07 & 0.03 &  0.09 & 0.04 &  0.10 & 0.06 &  0.45 & 0.24 &  0.63 & 0.13 & F5II        & 0.61 \\
27 & G284.2980--00.5951 & 10230784-5759512 &  7.49 & 0.02 &  1.44 & 0.03 &  0.58 & 0.04 &  0.38 & 0.06 &  0.44 & 0.03 &  0.70 & 0.03 &  0.90 & 0.03 &  1.24 & 0.02 & B           & 1.15 \\
28 & G283.9831--00.5360 & 10212249-5746404 & 13.39 & 0.04 &  1.80 & 0.44 &  0.54 & 0.05 &  0.21 & 0.06 &  0.15 & 0.05 &  0.12 & 0.08 &  \nodata & 1.00 &  1.62 & 0.20 & G2V         & 1.60 \\
29 & G284.0107--00.1372 & 10230932-5727250 & 14.05 & 0.05 &  2.08 & 0.45 &  0.48 & 0.05 &  0.11 & 0.05 &  0.11 & 0.07 &  0.06 & 0.09 &  \nodata & 1.00 &  1.58 & 0.20 & F8V         & 1.53 \\
30 & G283.9764--00.1365 & 10225656-5726168 & 13.36 & 0.04 &  0.63 & 0.02 &  0.63 & 0.05 &  0.23 & 0.05 &  0.30 & 0.06 &  0.36 & 0.07 &  \nodata & 1.00 &  1.09 & 0.14 & A7V         & 1.11 \\
31 & G284.4730--00.2456 & 10253880-5747417 & 12.33 & 0.03 &  0.65 & 0.43 &  0.18 & 0.04 &  0.03 & 0.04 &  0.06 & 0.04 &  0.09 & 0.05 &  0.23 & 0.11 &  0.34 & 0.07 & A7V         & 0.36 \\
32 & G283.9309--00.0712 & 10225509-5721302 & 12.58 & 0.03 &  1.91 & 0.44 &  0.53 & 0.04 &  0.11 & 0.17 &  0.04 & 0.04 &  0.01 & 0.05 &  0.24 & 0.10 &  0.44 & 0.11 & G0V         & 0.40 \\
33 & G283.9809--00.1931 & 10224464-5730119 & 12.78 & 0.02 &  \nodata & \nodata &  0.43 & 0.04 &  0.11 & 0.04 &  0.12 & 0.04 &  0.05 & 0.05 &  0.12 & 0.17 &  0.52 & 0.21 & F8V         & 0.52 \\
\enddata
\tablerefs{ (1) Reference ID \# for this paper; (2) GLIMPSE
Catalog ID; (3) 2MASS Catalog ID; (4) 2MASS K magnitude; (5) 1
$\sigma$ uncertainty for K magnitude; (6) Guide Star Catalog 2.2 V
- 2MASS K; (7) $1\sigma$ uncertainty for V-K; (8) 2MASS J - 2MASS
K; (9) $1\sigma$ uncertainty for J-K; (10) 2MASS H - 2MASS K;
(11) $1\sigma$ uncertainty for H-K; (12) 2MASS K - IRAC [3.6]; (13) $1\sigma$
uncertainty for K-[3.6]; (14) 2MASS K - IRAC [4.5]; (15) $1\sigma$ uncertainty
for K-[4.5]; (16) 2MASS K - IRAC [5.8]; (17) $1\sigma$ uncertainty for K-[5.8]; (18)
2MASS K - IRAC [8.0]; (19) $1\sigma$ uncertainty for K-[8.0];
(20) Spectral type of star; (21)E($K-8$) for star.}
\end{deluxetable}

\begin{deluxetable}{ccccccccccc}
\rotate
\tabletypesize{\tiny}
\setlength{\tabcolsep}{0.01in}
\tablewidth{6.0in}
\tablecaption{Derived Parameters for Candidate
$\beta$ Pictoris Analogs}
\tablehead{
\colhead{\ }&
\colhead{ID} &
\colhead{2MASS ID} &
\colhead{Spec.} &
\colhead{E(K-8)} &
\colhead{Distance} &
\colhead{A$_{v}$} &
\colhead{Temp} &
\colhead{$\frac{L_{IR}}{L_{*}}$} &
\colhead{Teff} &
\colhead{Log g} \\
\colhead{\ } &
\colhead{} &
\colhead{} &
\colhead{} &
\colhead{[mag]} &
\colhead{[pc]} &
\colhead{[mag]} &
\colhead{[K]} &
\colhead{} &
\colhead{[K]} &
\colhead{}\\
\colhead{1} &
\colhead{2} &
\colhead{3} &
\colhead{4} &
\colhead{5} &
\colhead{6} &
\colhead{7} &
\colhead{8} &
\colhead{9} &
\colhead{10} &
\colhead{11}  }
\startdata
 1 & G284.3535--00.2021 & 10250358-5741409 & A3V         & 1.00 & 1068$^{+ 160}_{- 160}$  & 0.00$^{+ 0.07}_{- 0.00}$ & 385$^{+ 70}_{- 56}$   & 0.0043$^{+ 0.0014}_{- 0.0010}$ & 8200  & +4.29 \\
 2 & G284.1744--00.5141 & 10224039-5751461 & A2V         & 0.92 & 1398$^{+ 210}_{- 210}$  & 0.00$^{+ 0.08}_{- 0.00}$ & 300$^{+ 37}_{- 30}$   & 0.0059$^{+ 0.0025}_{- 0.0014}$ & 8200  & +4.29 \\
 3 & G283.8842--00.3361 & 10213378-5733229 & B9V         & 0.77 & 1938$^{+ 290}_{- 290}$  & 0.00$^{+ 0.08}_{- 0.00}$ & 482$^{+ 113}_{- 95}$  & 0.0018$^{+ 0.0004}_{- 0.0002}$ & 9520  & +4.14 \\
 4 & G284.1241--00.2429 & 10232673-5736249 & A7V         & 0.54 & 563$^{+ 85}_{- 85}$     & 0.23$^{+ 0.09}_{- 0.08}$ & 620$^{+ 129}_{- 129}$ & 0.0033$^{+ 0.0005}_{- 0.0004}$ & 7200  & +4.34 \\
 5 & G284.0185--00.1803 & 10230188-5729511 & K0V         & 0.43 & 593$^{+ 90}_{- 90}$     & 0.17$^{+ 0.19}_{- 0.17}$ & 801$^{+ 74}_{- 56}$   & 0.0078$^{+ 0.0012}_{- 0.0010}$ & 5250  & +4.49 \\
 6 & G284.0547--00.5695 & 10214145-5750415 & B8V         & 1.24 & 3073$^{+ 475}_{- 460}$  & 1.47$^{+ 0.19}_{- 0.16}$ & 350$^{+ 42}_{- 46}$   & 0.0137$^{+ 0.0076}_{- 0.0030}$ & 11900 & +4.04 \\
 7 & G284.0719--00.1637 & 10232641-5730435 & F3IV        & 0.62 & 963$^{+ 145}_{- 150}$   & 0.78$^{+ 0.32}_{- 0.20}$ & 803$^{+ 95}_{- 149}$  & 0.0081$^{+ 0.0011}_{- 0.0009}$ & 6440  & +4.34 \\
 8 & G284.0110--00.1208 & 10231333-5726356 & F5IV        & 0.46 & 918$^{+ 140}_{- 140}$   & 0.62$^{+ 0.25}_{- 0.23}$ & 838$^{+ 181}_{- 240}$ & 0.0057$^{+ 0.0009}_{- 0.0009}$ & 6440  & +4.34 \\
 9 & G284.2320--00.1670 & 10242577-5736016 & A5V         & 0.61 & 773$^{+ 115}_{- 115}$   & 0.41$^{+ 0.21}_{- 0.21}$ & 508$^{+ 110}_{- 73}$  & 0.0030$^{+ 0.0005}_{- 0.0005}$ & 8200  & +4.29 \\
10 & G284.0658--00.3254 & 10224485-5738431 & G0V         & 0.55 & 493$^{+ 75}_{- 75}$     & 0.04$^{+ 0.23}_{- 0.04}$ & 567$^{+ 106}_{- 85}$  & 0.0051$^{+ 0.0007}_{- 0.0009}$ & 6030  & +4.49 \\
11 & G283.9773--00.3948 & 10215463-5739217 & B8V         & 0.42 & 2798$^{+ 420}_{- 420}$  & 0.41$^{+ 0.17}_{- 0.17}$ & 460$^{+ 282}_{- 102}$ & 0.0009$^{+ 0.0002}_{- 0.0002}$ & 11900 & +4.04 \\
12 & G283.9935--00.1944 & 10224911-5729455 & G5V         & 1.03 & 553$^{+ 85}_{- 85}$     & 0.27$^{+ 0.29}_{- 0.09}$ & 424$^{+ 50}_{- 44}$   & 0.0154$^{+ 0.0025}_{- 0.0025}$ & 5770  & +4.49 \\
13 & G283.9239--00.5103 & 10210640-5743271 & F5IV/V      & 0.71 & 793$^{+ 120}_{- 120}$   & 0.00$^{+ 0.29}_{- 0.00}$ & 542$^{+ 103}_{- 81}$  & 0.0050$^{+ 0.0009}_{- 0.0005}$ & 6440  & +4.34 \\
14 & G283.9153--00.4337 & 10212181-5739183 & F3V         & 0.93 & 1068$^{+ 165}_{- 175}$  & 0.39$^{+ 0.13}_{- 0.24}$ & 395$^{+ 54}_{- 34}$   & 0.0115$^{+ 0.0027}_{- 0.0017}$ & 6440  & +4.34 \\
15 & G284.0478--00.1686 & 10231576-5730119 & K5V         & 0.91 & 608$^{+ 95}_{- 90}$     & \nodata                  & 654$^{+ 60}_{- 54}$   & 0.0309$^{+ 0.0022}_{- 0.0026}$ & 4350 & +4.54 \\
16 & G283.9076--00.1997 & 10221550-5727151 & G8IV        & 0.49 & 268$^{+ 40}_{- 40}$     & \nodata                  & 723$^{+ 52}_{- 67}$   & 0.0066$^{+ 0.0010}_{- 0.0007}$ & 5250 & +4.49 \\
17 & G283.9040--00.3687 & 10213333-5735398 & K5V         & 0.50 & 338$^{+ 50}_{- 50}$     & \nodata                  & 776$^{+ 33}_{- 74}$   & 0.0148$^{+ 0.0029}_{- 0.0073}$ & 4350 & +4.54 \\
18 & G284.3417--00.2049 & 10245840-5741272 & F8Ve        & 1.05 & 513$^{+ 75}_{- 75}$     & \nodata                  & 560$^{+ 84}_{- 43}$   & 0.0108$^{+ 0.0002}_{- 0.0002}$ & 6030 & +4.49 \\
\enddata

\tablerefs{ (1) Reference ID \# for this paper; (2) GLIMPSE Catalog
ID; (3) 2MASS Catalog ID; (4) Spectral type of star; (5) E($K-8$) for
star; (6) Spectrophotometric distance; (7) Visual extinction;
(8) Best fit blackbody model temperature for excess; (9) Best fit disk
to star luminosity ratio; (10) Effective temperature for the adopted
Kurucz model (11) Log g for the adopted Kurucz model}
\end{deluxetable}

\clearpage

\clearpage

\clearpage

\clearpage

\clearpage

\clearpage

\clearpage

\clearpage

\clearpage

\clearpage

\clearpage

\clearpage

\clearpage

\clearpage

\end{document}